\documentclass[prb,twocolumn,superscriptaddress,floatfix,longbibliography]{revtex4-1}
\usepackage[english]{babel}
\usepackage{graphicx}
\usepackage{amsmath}
\usepackage{amssymb}
\usepackage{color}
\usepackage[normalem]{ulem}
\usepackage{hyperref}
\newcommand{\cb}{\mathrm{c}}
\newcommand{\vb}{\mathrm{v}}
\newcommand{\mrd}{\mbox{\hspace*{0.6mm}}}
\begin{document}
\title{Identification of Ultrafast Photophysical Pathways in Photoexcited Organic Heterojunctions}
\author{Veljko Jankovi\'c} \email{veljko.jankovic@ipb.ac.rs}
\affiliation{Scientific Computing Laboratory, Center for the Study of Complex
Systems, Institute of Physics Belgrade, University of Belgrade, 
Pregrevica 118, 11080 Belgrade, Serbia}
\author{Nenad Vukmirovi\'c} \email{nenad.vukmirovic@ipb.ac.rs}
\affiliation{Scientific Computing Laboratory, Center for the Study of Complex
Systems, Institute of Physics Belgrade, University of Belgrade, 
Pregrevica 118, 11080 Belgrade, Serbia}

\begin{abstract} 
The exciton dissociation and charge separation occurring on subpicosecond time scales following the photoexcitation are
studied in a model donor/acceptor heterojunction using a fully quantum approach.
Higher-than-LUMO acceptor orbitals which are energetically aligned with the donor LUMO orbital participate in the ultrafast interfacial dynamics
by creating photon-absorbing charge-bridging states in which charges are spatially separated and which can be directly photoexcited.
Along with the states brought about by single-particle resonances,
the two-particle (exciton) mixing gives rise to bridge states in which charges are delocalized.
Bridge states open up a number of photophysical pathways that indirectly connect the initial donor states with states of spatially separated charges
and compete with the efficient progressive deexcitation within the manifold of donor states.
The diversity and efficiency of these photophysical pathways depend on a number of factors, such as the precise energy alignment of exciton states,
the central frequency of the excitation, and the strength of carrier-phonon interaction.
\end{abstract}

\maketitle{}

\section{Introduction}
Tremendous research efforts have been devoted to understanding the microscopic mechanisms governing efficient and ultrafast (happening on a $\lesssim 100$-fs time scale)
free-charge generation observed in time-resolved experiments on
donor/acceptor (D/A) heterojunction organic photovoltaic (OPV) devices.~\cite{nmat12-29,nmat12-66,science343-512,JAmChemSoc.136.1472}
The photogenerated exciton in the donor material is commonly believed to transform into the charge transfer (CT) exciton.~\cite{ChemRev.110.6736,ADMA:ADMA201000376}
In the CT exciton, the electron and hole are tightly bound and localized at the D/A interface.
The Coulomb barrier preventing the electron and hole in the CT state from further charge separation and formation of a charge separated (CS)
state is much higher than the thermal energy at room temperature, so that the actual mechanism of the emergence
of spatially separated charges on such short time scales remains an open question.~\cite{PCCP.16.20291,PCCP.17.2311,PCCP.17.28451,ChemRev.116.13279}

Electronically hot CT states,
which are essentially resonant with the initial states of donor excitons
and exhibit significant charge delocalization,~\cite{FD.163.377,PhysRevB.88.205304}
are believed to be precursors to separated charges present on ultrafast time scales following the excitation.~\cite{nmat12-29,nmat12-66,science335-1340,JAmChemSoc.136.2876,JPhysChemC.119.15028,jacs.135.18502}
The delocalization of carriers can also reduce the Coulomb barrier and allow the transition from CT to CS exciton.~\cite{science343-512,JAmChemSoc.135.16364,PhysChemChemPhys.16.20305,PhysRevB.90.115420}
The ultrafast exciton dissociation and charge separation are not purely electronic processes,
but are instead mediated by the carrier-phonon coupling.~\cite{PhysRevLett.100.107402,JPhysChemC.115.10205,JChemPhys.137.22A540,JAmChemSoc.135.16364,JPhysChemLett.6.1702,ncomms5-3119,PhysRevB.91.201302,Science.344.1001}
The phonon-mediated ultrafast exciton dissociation and charge separation
can proceed via the so called intermediate bridge states,~\cite{PhysRevLett.100.107402,PhysRevB.91.201302}
the vibronically hot CT states,~\cite{JAmChemSoc.135.16364} or can occur without any intermediate CT state.~\cite{ncomms5-3119}
The exciton states of mixed donor and CS character are found to open up different photophysical pathways for
ultrafast dissociation of initial donor excitons,
which are concurrent with vibronically-assisted transitions within the donor exciton manifold.~\cite{JPhysChemLett.6.1702}

We have recently investigated the exciton dynamics occurring on a subpicosecond
time scale following the excitation of the model D/A heterojunction.~\cite{PhysRevB.95.075308}
Our model explicitly takes into account the physical mechanisms regarded as highly relevant for the ultrafast heterojunction dynamics,
such as the carrier delocalization and the carrier-phonon interaction.
Moreover, the exciton generation, exciton dissociation, and further charge separation are treated
on equal footing and on a fully quantum level, which is essential to correctly describe processes taking place on ultrafast time scales.
For the model parameters representative of a low-bandgap polymer/fullerene blend,
we found that the major part of space-separated charges present on ~100-fs time scales after the excitation originates
from the direct optical generation from the ground state rather than from the ultrafast population transfer from initially generated donor
excitons. The resonant mixing between single-electron states in the two materials leads to the redistribution of oscillator strengths between
states of donor excitons and space-separated charges, the latter becoming accessible by direct photoexcitation.

In this study, we aim at giving a more detailed description of the ultrafast heterojunction dynamics
in terms of particular photophysical pathways along which it proceeds.
In order to keep the numerical effort within reasonable limits, we still use one-dimensional model of a heterojunction,
but we extend it by taking into account more than only one single-electron (single-hole) state per site.
The model parameters are chosen to be representative of the prototypical blend of
poly-3-hexylthiophene (P3HT) and [6,6]-phenyl-C$_{61}$ butyric acid methyl ester (PCBM).
The aforementioned extension of the model is important in many aspects.
First, the degeneracy of the LUMO, LUMO+1, and LUMO+2 orbitals of the C$_{60}$ molecule is broken in its functionalized derivative PCBM,~\cite{NanoLett.7.1967,ChinPhysB.21.017102,JMaterChemC.2.7313}
giving rise to three energetically close bands of electronic states of PCBM aggregates.
This fact was shown to be important for efficient and ultrafast charge separation observed in D/A blends containing PCBM as the acceptor.~\cite{PhysRevB.91.201302,JAmChemSoc.136.2876,AdvMater.25.1038}
Upon the functionalization of C$_{60}$, together with the degeneracy of its LUMO, LUMO+1, and LUMO+2 orbitals,
the degeneracy of its LUMO+3, LUMO+4, and LUMO+5 orbitals, which are situated at around 1 eV above the LUMO, LUMO+1, and LUMO+2 orbitals,
is also broken.
Second, according to the results of Ma and Troisi~\cite{JPhysChemC.118.27272},
the precise energy alignment of higher-than-LUMO orbitals of the acceptor can modulate the exciton dissociation rate by orders
of magnitude by opening up new exciton dissociation channels.
The LUMO-LUMO offset in the P3HT/PCBM blend can be quite large (around 1 eV)~\cite{JPhysChemLett.2.1099,JMaterChem.21.1479,JPhysChemC.115.2406}
and thus comparable to the energy separation between LUMO and LUMO+3 orbitals of the PCBM molecule.
It can therefore be expected that the electronic states of a PCBM aggregate which arise
from LUMO+3, LUMO+4, and LUMO+5 orbitals of the PCBM molecule may play nontrivial role in
the ultrafast interfacial dynamics.
Surprisingly, it seems that the effect of these orbitals
has not received enough attention in previous model studies of the P3HT/PCBM heterojunction. 
The ultrafast electron transfer observed in ref~\citenum{ApplPhysLett.98.083303} has been ascribed to
the energy overlap between the state of the photoexcited electron and the electronic states of the fullerene aggregate.  
The result presented in Figure 3e of ref~\citenum{ApplPhysLett.98.083303} suggests that this overlap involves the electronic
states of the fullerene aggregate stemming from the LUMO+3, LUMO+4, and LUMO+5 orbitals of the PCBM molecule.

Our results indicate that the exciton states in which the charges are delocalized throughout the heterojunction
play a crucial role in the ultrafast heterojunction dynamics.
In the low-energy part of the exciton spectrum,
such states emerge due to the resonant mixing between different exciton (i.e., two-particle) states and we denote them as bridge states.
However, in the high-energy region of the exciton spectrum,
such states form as a consequence of the resonant mixing between single-electron states in the donor and acceptor
(states originating from LUMO+3, LUMO+4, and LUMO+5 orbitals of the PCBM molecule).
The relevant exciton states of this kind are those in which the charges are spatially separated
(the electron is mainly in the acceptor, while the hole is mainly in the donor) and we denote them as
photon-absorbing charge-bridging (PACB) states,~\cite{NanoLett.7.1967,PhysChemChemPhys.13.21461,PhysChemChemPhys.18.9514}
since they can be directly reached by a photoexcitation.
Exciting well above the lowest donor state, we find that excitons are generated in both donor and PACB states,
while the major part of space-separated charges present on a 100-fs time scale following the excitation resides in PACB states.
The deexcitation of initial PACB excitons proceeds via the donor exciton manifold,
while single-phonon-assisted processes involving a PACB state and
CT and CS states belonging to the low-energy part of the spectrum are virtually absent.
The donor excitons mainly deexcite within the donor exciton manifold and,
before reaching the lowest donor exciton state, may perform transitions to bridge states,
which are gateways into the space-separated manifold.
The lowest donor state, being essentially decoupled from the space-separated manifold,
is a trap state for exciton dissociation.
The bridge states can be either intermediate or final states in the course of the charge separation.
Once a space-separated state is reached, the gradual energy loss within the space-separated manifold leads
to the population of low-energy CT states on a picosecond time scale.
The participation of PACB excitons in the total exciton population strongly depends on the central frequency of the excitation.
The probability of a bridge state being accessed during the exciton deexcitation sensitively
depends on the distribution of initially generated excitons, the energy level alignment, and the carrier-phonon interaction strength.

\section{Model and Methods}
\subsection{Model Hamiltonian}
We use the standard semiconductor Hamiltonian with multiple single-electron/single-hole states per site.
The model heterojunction consists of $2N$ sites located on a one-dimensional lattice of constant $a$:
sites $0,\dots,N-1$ belong to the donor part,
while sites $N,\dots,2N-1$ belong to the acceptor part of the heterojunction.
The single-electron levels on site $i$ are counted by index $\beta_i$,
so that Fermi operators $c_{i\beta_i}^\dagger$ ($c_{i\beta_i}$)
create (destroy) electrons on site $i$ and in single-electron state $\beta_i$.
Analogously, single-hole levels on site $i$ are counted by index $\alpha_i$,
so that Fermi operators $d_{i\alpha_i}^\dagger$ ($d_{i\alpha_i}$)
create (destroy) holes on site $i$ and in single-hole state $\alpha_i$.
Each site contributes a number of localized phonon modes and the corresponding Bose operators
$b_{i\lambda_i}^\dagger$ ($b_{i\lambda_i}$) create (annihilate) phonons
belonging to mode $\lambda_i$ on site $i$.
The Hamiltonian has the form
\begin{equation}
\label{Eq:H_tot}
 H=H_\mathrm{c}+H_\mathrm{p}+H_\mathrm{c-p}+H_\mathrm{c-f},
\end{equation}
where $H_\mathrm{c}$ describes interacting carriers
\begin{equation}
\label{Eq:H_c}
\begin{split}
 H_\mathrm{c}&=\sum_{\substack{i\beta_i\\j\beta'_j}}\epsilon^\cb_{(i\beta_i)(j\beta'_j)} c_{i\beta_i}^\dagger c_{j\beta'_j}
-\sum_{\substack{i\alpha_i\\j\alpha'_j}}\epsilon^\vb_{(i\alpha_i)(j\alpha'_j)} d_{i\alpha_i}^\dagger d_{j\alpha'_j}\\
&+\frac{1}{2}\sum_{\substack{i\beta_i\\j\beta'_j}}V_{ij}\mrd c_{i\beta_i}^\dagger c_{j\beta'_j}^\dagger c_{j\beta'_j} c_{i\beta_i}\\
&+\frac{1}{2}\sum_{\substack{i\alpha_i\\j\alpha'_j}}V_{ij}\mrd d_{i\alpha_i}^\dagger d_{j\alpha'_j}^\dagger d_{j\alpha'_j} d_{i\alpha_i}\\
&-\sum_{\substack{i\beta_i\\j\alpha_j}}V_{ij}\mrd c_{i\beta_i}^\dagger d_{j\alpha_j}^\dagger d_{j\alpha_j} c_{i\beta_i},
\end{split}
\end{equation}
\begin{equation}
\label{Eq:H_p}
 H_\mathrm{p}=\sum_{\substack{i\lambda_i}}\hbar\omega_{i\lambda_i} b_{i\lambda_i}^\dagger b_{i\lambda_i}
\end{equation}
is the phonon part of the Hamiltonian, $H_\mathrm{c-p}$ accounts for the carrier-phonon interaction
\begin{equation}
\label{Eq:H_c_p}
\begin{split}
 H_\mathrm{c-p}&=\sum_{i\beta_i}\sum_{\lambda_i} g^\mathrm{c}_{i\beta_i\lambda_i} c_{i\beta_i}^\dagger c_{i\beta_i}(b_{i\lambda_i}^\dagger+b_{i\lambda_i})\\
&-\sum_{i\alpha_i}\sum_{\lambda_i} g^\mathrm{v}_{i\alpha_i\lambda_i} d_{i\alpha_i}^\dagger d_{i\alpha_i}(b_{i\lambda_i}^\dagger+b_{i\lambda_i}),
\end{split}
\end{equation}
whereas $H_\mathrm{c-f}$ represents the interaction of carriers with the external electric field $E(t)$
\begin{equation}
\label{Eq:H_c_f}
 H_\mathrm{c-f}=-E(t)\sum_{i\alpha_i\beta_i} d_{i\alpha_i\beta_i}^{\cb\vb}(c_{i\beta_i}^\dagger d_{i\alpha_i}^\dagger+d_{i\alpha_i} c_{i\beta_i}).
\end{equation}
In our model, quantities $\epsilon^\cb_{(i\beta_i)(j\beta'_j)}$ ($\epsilon^\vb_{(i\alpha_i)(j\alpha'_j)}$),
which represent electron (hole) on-site energies and transfer integrals, are nonzero only for certain combinations of their indices.
Namely, $\epsilon^\cb_{(i\beta_i)(j\beta'_j)}$ is non-zero when it represents
\begin{enumerate}
 \item on-site energy $\epsilon^\cb_{i\beta_i}$ of electron level $\beta_i$ on site $i$ for $i=j$ and $\beta_i=\beta'_i$;
 \item negative electron transfer integral between nearest neighbors of band $\beta_i$, $-J^\mathrm{c,int}_{i\beta_i}$,
for $i$ and $j$ both belonging to the same part of the heterojunction, $|i-j|=1$, and $\beta_i=\beta'_j$;
 \item negative electron transfer integral between nearest neighbors of different bands,
$-J^\mathrm{c,ext}_{i\beta_i\beta'_j}$, for $i$ and $j$ both belonging to the same part of the heterojunction, $|i-j|=1$, and $\beta_i\neq\beta'_j$;
 \item negative electron transfer integral between different parts of the heterojunctions, $-J^\cb_{DA}$, for $i=N-1$ and $j=N$ or vice versa.
\end{enumerate}
The Coulomb interaction described by eq~\ref{Eq:H_c} is taken into account in the lowest monopole-monopole approximation
and the interaction potential $V_{ij}$ is assumed to be the Ohno potential
\begin{equation}
\label{Eq:Ohno}
 V_{ij}=\frac{U}{\sqrt{1+\left(\frac{r_{ij}}{r_0}\right)^2}},
\end{equation}
where $U$ is the on-site Coulomb interaction,
$r_{ij}$ is the distance between sites $i$ and $j$,
$r_0=e^2/(4\pi\varepsilon_0\varepsilon_r U)$ is the characteristic length,
and $\varepsilon_r$ is the relative dielectric constant.
Charge carriers are assumed to be locally and linearly coupled to the set of phonon modes
(Holstein-type interaction), as given in eq~\ref{Eq:H_c_p}.
We assume that the frequency of the external electric field is such that it creates electron-hole excitations,
the interband matrix elements of the dipole moment being $d^{\cb\vb}_{i\alpha_i\beta_i}$,
and neglect all intraband dipole matrix elements.

\subsection{Theoretical Framework}
Ultrafast exciton dynamics governed by the model Hamiltonian defined in eqs~\ref{Eq:H_tot}-\ref{Eq:H_c_f}
is treated using the density matrix formalism complemented with the dynamics controlled truncation (DCT)
scheme.~\cite{ZPhysB.93.195,PhysRevB.53.7244,RevModPhys.70.145,PhysRevB.92.235208,PhysRevB.95.075308}
Exciton generation (from initially unexcited heterojunction) by means of a pulsed excitation
and subsequent evolution of thus created nonequilibrium
state of the system are treated on equal footing.
We consider the case of weak excitation and low carrier densities.
The carrier branch of the hierarchy of equations produced by the density matrix formalism
can then be truncated retaining only contributions up to the second order in the exciting field.
The truncation of the phonon branch of the hierarchy is performed to ensure the conservation
of the particle-number and energy after the pulsed excitation.~\cite{PhysRevB.92.235208}

It is advantageous to formulate theory in the subspace of single electron-hole excitations,
which is spanned by the so-called exciton basis.
The most general electron-hole pair state is of the form
$\displaystyle{|x\rangle=\sum_{\substack{i\alpha_i\\j\beta_j}}\psi^x_{(i\alpha_i)(j\beta_j)}c_{j\beta_j}^\dagger d_{i\alpha_i}^\dagger|0\rangle}$,
where $|0\rangle$ is the vacuum of electron-hole pairs.
The exciton basis states are obtained by solving the eigenvalue problem
$H_\mathrm{c}|x\rangle=\hbar\omega_x|x\rangle$, which in the basis of single-particle states localized at lattice sites reads as
\begin{equation}
\label{Eq:exc_eigen_prob}
\begin{split}
 \sum_{\substack{i'\alpha'_i\\j'\beta'_j}}\left(
\delta_{ii'}\delta_{\alpha_i\alpha'_i}\epsilon^\cb_{(j\beta_j)(j'\beta'_j)}
-\delta_{jj'}\delta_{\beta_j\beta'_j}\epsilon^\vb_{(i\alpha_i)(i'\alpha'_i)} \right. \\
- \left. \delta_{ii'}\delta_{\alpha_i\alpha'_i}\delta_{jj'}\delta_{\beta_j\beta'_j} V_{ij}
\right)\psi^x_{(i'\alpha'_i)(j'\beta'_j)}=\hbar\omega_x\mrd\psi^x_{(i\alpha_i)(j\beta_j)}.
\end{split}
\end{equation}
The operator which creates an exciton in state $x$ is defined through
\begin{equation}
 X_x^\dagger=\sum_{\substack{i\alpha_i\\j\beta_j}}\psi^x_{(i\alpha_i)(j\beta_j)} c_{j\beta_j}^\dagger d_{i\alpha_i}^\dagger.
\end{equation} 
The total Hamiltonian (eq~\ref{Eq:H_tot}), in which we keep only operators whose expectation values are at most
of the second order in the exciting field, can be expressed in terms of exciton creation and annihilation operators as
\begin{equation}
\label{Eq:exc_ham_2nd}
\begin{split}
 H=&\sum_x \hbar\omega_x X_x^\dagger X_x+\sum_{i\lambda_i}
\hbar\omega_{i\lambda_i} b_{i\lambda_i}^\dagger b_{i\lambda_i}\\
  +&\sum_{\substack{\bar x x\\i\lambda_i}}\left(\Gamma^{i\lambda_i}_{\bar x
x}X_{\bar x}^\dagger X_x b_{i\lambda_i}^\dagger+
\Gamma^{i\lambda_i*}_{\bar x x}X_{x}^\dagger X_{\bar x} b_{i\lambda_i}\right)\\
-&E(t)\sum_x\left(M_x^* X_x+M_x X_x^\dagger\right).
\end{split}
\end{equation}
Dipole-moment matrix elements for the direct generation (from the ground state)
of excitons in state $x$ are given as
\begin{equation}
\label{Eq:M_x}
 M_x=\sum_{i}\sum_{\alpha_i\beta_i}\psi^{x*}_{(i\alpha_i)(i\beta_i)} d^\mathrm{cv}_{i\alpha_i\beta_i},
\end{equation}
while exciton-phonon matrix elements describing transitions
from exciton state $x$ to exciton state $\bar x$ assisted by phonon $(i\lambda_i)$ are
\begin{equation}
\label{Eq:gamma}
\begin{split}
 \Gamma^{i\lambda_i}_{\bar x x}&=\sum_{\beta_i}\sum_{j\alpha_j} g^\cb_{i\beta_i\lambda_i}
\psi^{\bar x*}_{(j\alpha_j)(i\beta_i)}\psi^x_{(j\alpha_j)(i\beta_i)}\\
&-\sum_{\alpha_i}\sum_{j\beta_j} g^\vb_{i\alpha_i\lambda_i}
\psi^{\bar x*}_{(i\alpha_i)(j\beta_j)}\psi^x_{(i\alpha_i)(j\beta_j)}.
\end{split}
\end{equation}
Active variables in our formalism are electronic density matrices $y_x=\langle X_x\rangle$
and $n_{\bar x x}=\langle X_{\bar x}^\dagger X_x\rangle$, along with their single-phonon-assisted
counterparts $y_{x(i\lambda_i)^-}=\langle X_x
b_{i\lambda_i}\rangle$, $y_{x(i\lambda_i)^+}=\langle X_x
b_{i\lambda_i}^\dagger\rangle$, and
$n_{\bar x x(i\lambda_i)^+}=\langle X_{\bar x}^\dagger X_x
b_{i\lambda_i}^\dagger\rangle$.
The equations of motion for these variables are given in Supporting Information.
Since the phonon branch of the hierarchy is truncated at the level of second-order phonon assistance,
our treatment of the electron-phonon interaction does not capture properly the processes with higher-order phonon assistance,
which are important for stronger electron-phonon interaction.
In this case, the feedback effects of electronic excitations on phonons would have to be taken into account as well.
To this end, in our recent publication~\cite{PhysRevB.95.075308} we performed a computation of subpicosecond dynamics using surface hopping approach
(which, however, treats lattice dynamics classically)
and found that the feedback effects were not very pronounced. In order to treat the electron-phonon interaction more accurately,
other approaches based on state-of-the-art multiconfigurational techniques~\cite{JPhysChemLett.6.1702},
infinite resummation within Green's function formalism~\cite{PhysRevB.57.R2061,ChemPhysLett.677.87}
or variational ans\"{a}tze for the wave function of electron-phonon system~\cite{PhysRevB.91.041107}
can be used.

The early stages of our numerical experiment (during and immediately after the pulsed excitation)
are dominated by exciton coherences with the ground state
$y_x$ and their phonon-assisted counterparts.
The corresponding coherent exciton populations $|y_x|^2$ are not a measure of the number of
truly bound electron-hole pairs and generally decay quickly after the pulsed excitation,
converting into incoherent exciton populations. This conversion from coherent to incoherent
exciton populations is in our model mediated by the carrier-phonon interaction.
The incoherent exciton populations are defined as
\begin{equation}
\label{Eq:incoh_coh_tot_x}
 \bar n_{xx}=n_{xx}-|y_x|^2.
\end{equation}
They represent numbers of Coulomb-correlated electron-hole
pairs and typically exist for a long time after the decay of coherent populations.
The incoherent populations of various groups $X$ of exciton states are defined as
\begin{equation}
 N_X^\mathrm{incoh}=\sum_{x\in X}\bar n_{xx},
\end{equation}
and are frequently and conveniently normalized to the total exction population
\begin{equation}
\label{Eq:pop_total}
 N_\mathrm{tot}=\sum_x n_{xx},
\end{equation}
which is conserved after the excitation.
Once created from coherent populations, incoherent populations redistribute among
various exciton states, the redistribution being mediated by the carrier-phonon interaction.
In order to gain insight into the pathways along which these redistribution processes
proceed, we define
energy- and time-resolved exciton populations
$\varphi_X(E,t)$ of states belonging to group $X$ as
\begin{equation}
\label{Eq:ene_res_tot}
 \varphi_X(E,t)=\frac{1}{N_\mathrm{tot}}\sum_{x\in X} n_{xx}(t)\mrd\delta(E-\hbar\omega_x),
\end{equation}
so that $\varphi_X(E,t)\Delta E$ is the number (normalized to $N_\mathrm{tot}$) of excitons from group $X$ 
residing in the states whose energies are between $E$ and $E+\Delta E$. Bearing in mind eq~\ref{Eq:incoh_coh_tot_x},
relating the coherent, incoherent, and total exciton population of state $x$,
quantity $\varphi_X(E,t)$ can be decomposed into its coherent
\begin{equation}
\label{Eq:ene_res_coh}
 \varphi_X^\mathrm{coh}(E,t)=\frac{1}{N_\mathrm{tot}}\sum_{x\in X} |y_x(t)|^2\mrd\delta(E-\hbar\omega_x),
\end{equation}
and incoherent part
\begin{equation}
\label{Eq:ene_res_incoh}
 \varphi_X^\mathrm{incoh}(E,t)=\frac{1}{N_\mathrm{tot}}\sum_{x\in X} \bar n_{xx}(t)\mrd\delta(E-\hbar\omega_x).
\end{equation}
The plots of $\varphi_X^\mathrm{coh}$ as a function of $E$ and $t$ provide information about states in which excitons
are initially generated (the initial exciton distribution)
and the time scale on which the conversion from coherent to incoherent exciton populations
takes place.
The plots of $\varphi_X^\mathrm{incoh}$ as a function of $E$ and $t$ reveal actual pathways along which
(incoherent) excitons are redistributed, starting from the initial exciton distribution.

\subsection{Parameterization of the Model Hamiltonian}
\begin{figure}[htbp]
 \centering
 \includegraphics{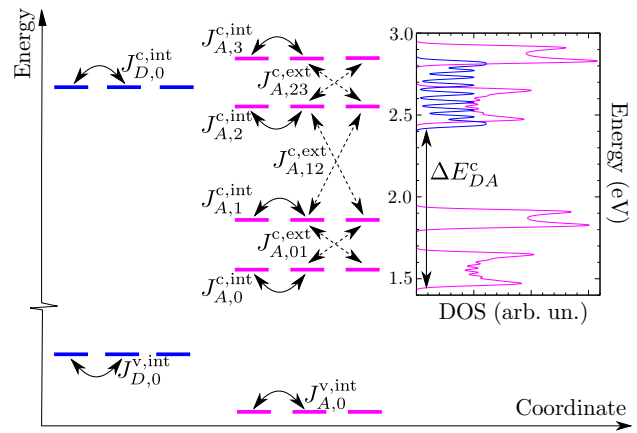}
 \caption{Illustration of the model system indicating different
transfer integrals present in Table~\ref{Tab:model_params_gen}.
The plot on the right shows the single-particle DOS for electrons
in the neat donor (blue curve) and acceptor (magenta curve) materials obtained using the values of
relevant parameters listed in Table~\ref{Tab:model_params_gen}.
The electronic states of the isolated materials are computed by diagonalizing
the free-electron Hamiltonian (the first term on the right-hand side of eq~\ref{Eq:H_c})
in which the D/A coupling is set to 0.
The DOS was then calculated by broadening each of the states obtained by a Gaussian with the standard deviation of 10 meV.}
 \label{Fig:fig_params}
\end{figure}

Our model is parametrized with the aim of describing ultrafast exciton dissociation and charge separation
in the direction perpendicular to the D/A interface. This is motivated by recent studies of ultrafast exciton dissociation~\cite{JChemPhys.138.164905}
and charge separation~\cite{PhysRevB.90.115420} in two-dimensional models of a D/A polymeric heterojunction which have suggested that these processes
crucially depend on the electronic properties and geometry in the direction perpendicular to the interface.
In actual computations, we take one single-electron level per site in the donor and
one single-hole level per site in both the donor and acceptor. In order to mimic the presence of
higher-than-LUMO orbitals energetically close to the LUMO level (which is a situation typical of fullerenes),
as well as to investigate the effects of single-electron levels situated at around 1.0 eV above the LUMO level on the exciton dissociation,
we take four single-electron levels per site in the acceptor.
Different types of electronic couplings are schematically indicated in Figure~\ref{Fig:fig_params},
while the values of model parameters used in computations are summarized in Table~\ref{Tab:model_params_gen}.
\begin{table}[htbp]
 \caption{Values of Model Parameters Used in Computations.}
 \label{Tab:model_params_gen}
 \centering
 \begin{tabular}{|c c|}
  \hline
  parameter & value\\
  \hline
  $N$ & 11\\
  $a$ (nm) & 1.0\\
  $U$ (eV) & 0.65\\
  $\varepsilon_r$ & 3.0\\
  $\epsilon^\cb_{D,0}$ (eV) & 2.63\\
  $J^{\cb,\mathrm{int}}_{D,0}$ (eV) & 0.1\\
  $\epsilon^\vb_{D,0}$ (eV) & $-$0.3\\
  $J^{\vb,\mathrm{int}}_{D,0}$ (eV) & $-$0.15\\
  $\epsilon^\cb_{A,0}$ (eV) & 1.565\\
  $\epsilon^\cb_{A,1}$ (eV) & 1.865\\
  $\epsilon^\cb_{A,2}$ (eV) & 2.565\\
  $\epsilon^\cb_{A,3}$ (eV) & 2.865\\
  $J^{\cb,\mathrm{int}}_{A,0}$ (eV) & 0.05\\
  $J^{\cb,\mathrm{int}}_{A,1}$ (eV) & 0.025\\
  $J^{\cb,\mathrm{int}}_{A,2}$ (eV) & 0.05\\
  $J^{\cb,\mathrm{int}}_{A,3}$ (eV) & 0.025\\
  $J^{\cb,\mathrm{ext}}_{A,01}$ (eV) & 0.02\\
  $J^{\cb,\mathrm{ext}}_{A,12}$ (eV) & 0.02\\
  $J^{\cb,\mathrm{ext}}_{A,23}$ (eV) & 0.02\\
  $\epsilon^\vb_{A,0}$ (eV) & $-$1.03\\
  $J^{\vb,\mathrm{int}}_{A,0}$ (eV) & $-$0.15\\ 
  $J^\cb_{DA}$ (eV) & 0.1\\
  $J^\vb_{DA}$ (eV) & $-$0.1\\
  $\hbar\omega_\mathrm{p,1}$ (meV) & 10.0\\
  $g_1$ (meV) & 42.0\\
  $\hbar\omega_\mathrm{p,2}$ (meV) & 185.0\\
  $g_2$ (meV) & 94.0\\
  $T$ (K) & 300.0\\
  \hline
 \end{tabular}
\end{table}
These values are selected so that the main characteristics of the single-particle and exciton spectrum
(band widths, band alignments, exciton and charge transfer state binding energies)
within the model correspond to the ones observed in P3HT/PCBM material system.
We take the HOMO level of the donor material to be the zero of the energy scale.

The value of the lattice spacing $a$ agrees with the typical distance between constitutive elements of organic semiconductors.
The number of sites in a single material $N=11$ is reasonable since
typical linear dimensions of phase segregated domains
in bulk heterojunction morphology are 10-20 nm.~\cite{afm22-1116}
The value of the transfer integral $J^{\vb,\mathrm{int}}_{D,0}$ was chosen so as to agree with
the HOMO bandwidth along the $\pi$-stacking direction
of the regioregular P3HT~\cite{PhysRevB.76.245202,PhysRevB.79.115207}
and the values of the hole transfer integral along the $\pi$-stacking direction
of the same material.~\cite{JPhysChemB.112.14857,JPhysChemB.113.9393}
The electron transfer integral $J^{\cb,\mathrm{int}}_{D,0}$
should be of similar magnitude as the hole transfer
integral along the $\pi$-stacking direction.~\cite{JPhysChemB.112.14857}
Energies of the single-electron and single-hole levels in the donor,
as well as the on-site Coulomb interaction $U$, were chosen
so that the lowest donor exciton state is located at around 2.0 eV, while
the HOMO-LUMO gap (single-particle gap) is around 2.4 eV,
i.e., the binding energy of the donor exciton is around 0.4 eV.~\cite{JPhysChemC.118.21873,Polymer.55.2667}

Electron transfer integrals in the acceptor
$J^{\cb,\mathrm{int}}_{A,0},J^{\cb,\mathrm{int}}_{A,1}$ and $J^{\cb,\mathrm{ext}}_{A,01}$,
together with the energy difference $\epsilon^\cb_{A,1}-\epsilon^\cb_{A,0}$ between single-electron states,
are chosen to reproduce the most important features of the low-energy part of the electronic density of states (DOS)
of fullerene aggregates,~\cite{JAmChemSoc.136.2876,PhysRevB.91.201302}
such as the combined (total) bandwidth of $0.4-0.5$ eV and the presence of two separated
groups of allowed states. Let us note that, because of the reduced dimensionality of our model,
we cannot expect to reproduce details of the actual DOS, but only its gross features.
We therefore believe that taking two instead of three orbitals energetically close to the LUMO orbital is
reasonable within our model. 
The electronic DOS in the acceptor produced by our model is shown in the inset of Figure~\ref{Fig:fig_params}.
Magnitudes of transfer integrals in the acceptor are also in agreement with the values
reported in the literature.~\cite{JPhysChemC.114.20479,PhysRevB.85.054301}
We have also included the single-electron fullerene states which are located at around 1 eV above the lowest single-electron state.
It is well known that these states in C$_{60}$ are also triply degenerate and that this degeneracy is lifted in PC$_{60}$BM.
Since we use a model system, we take, for simplicity, that the degeneracy is lifted in the same manner as in the case of lowest single-electron levels,
i.e., we take $J^{\cb,\mathrm{int}}_{A,0}=J^{\cb,\mathrm{int}}_{A,2}$, $J^{\cb,\mathrm{int}}_{A,1}=J^{\cb,\mathrm{int}}_{A,3}$,
$J^{\cb,\mathrm{int}}_{A,01}=J^{\cb,\mathrm{int}}_{A,23}$, and $\epsilon^\cb_{A,3}-\epsilon^\cb_{A,2}=\epsilon^\cb_{A,1}-\epsilon^\cb_{A,0}$,
while $\epsilon^\cb_{A,2}-\epsilon^\cb_{A,0}=1.0$ eV. For the magnitudes of the energy difference $\epsilon^\cb_{A,0}-\epsilon^\vb_{A,0}$
and the transfer integral $J^{\vb,\mathrm{int}}_{A,0}$ listed in Table~\ref{Tab:model_params_gen},
the single-particle gap in the acceptor part of the heterojunction assumes
the value of 2.2 eV, which is similar to the literature values for PCBM.~\cite{JPhysChemC.118.21873}
 
The energy differences $\Delta_\mathrm{XD-CT}$ and $\Delta_\mathrm{XA-CT}$
between the lowest excited state
of the heterojunction (the lowest CT state) and the lowest exciton states
in the donor and acceptor respectively, are directly related to LUMO-LUMO
and HOMO-HOMO energy offsets between the materials.
Literature values of $\Delta_\mathrm{XD-CT}$ representative of P3HT/PCBM blends
are usually calculated for the system consisting of one PCBM molecule and one oligomer
and range from 0.7 eV~\cite{JPhysChemLett.2.1099} to 1.3 eV.~\cite{JMaterChem.21.1479}
Liu and Troisi~\cite{JPhysChemC.115.2406} obtained $\Delta_\mathrm{XD-CT}=0.97$ eV
and pointed out that taking into account partial electron delocalization
over fullerene molecules can significantly lower the XD-CT energy difference.
For parameters listed in Table~\ref{Tab:model_params_gen},
$\Delta_\mathrm{XD-CT}=0.68$ eV, which is a reasonable value, since we do account for
carrier delocalization effects. The LUMO-LUMO offset $\Delta E^\mathrm{c}_{DA}$ (see Figure~\ref{Fig:fig_params})
produced by the model parameters
is around 0.96 eV and the lowest CT state is located at 1.32 eV.
The energy difference $\Delta_\mathrm{XA-CT}=0.42$ eV, so that the
HOMO-HOMO offset is around 0.73 eV and the lowest XA state is approximately at 1.74 meV,
both of which compare well with the available data.~\cite{JPhysChemC.118.21873}
The magnitudes of the transfer integrals $J^\cb_{DA}$ and $J^\vb_{DA}$ between the two materials
are taken to be similar to the values obtained in ref~\citenum{NanoLett.7.1967}.

Interband matrix elements of the dipole moment $d^{\cb\vb}_{i\alpha_i\beta_i}$ are assumed not to
depend on band indices $\alpha_i,\beta_i$ and to be equal on all sites belonging to the single
material, $d^{\cb\vb}_{i\alpha_i\beta_i}=d^{\cb\vb}_D$ for $i=0,\dots,N-1$ and
$d^{\cb\vb}_{i\alpha_i\beta_i}=d^{\cb\vb}_A$ for $i=N,\dots,2N-1$. Since the focus of our study is
on the dissociation of donor excitons, in all the computations we set $d^{\cb\vb}_A=0$.

We assume that each site contributes one low-frequency and one high-frequency phonon mode.
The energies of the phonon modes, as well as the carrier-phonon interaction constants, are taken to be
equal in both parts of the heterojunction.
The phonon mode of energy 185 meV, present in both materials, was shown to be important for ultrafast charge transfer in the P3HT/PCBM blend,~\cite{Science.344.1001}
while low-energy ($\lesssim 10$ meV) phonon modes of P3HT exhibit strong coupling to carriers.~\cite{JPhysChemB.120.5572} 
The strength of the carrier-phonon interaction can be quantified by the polaron binding energy,
which can be estimated using the result of the second-order
weak-coupling perturbation theory at $T=0$ in the vicinity of the point $k=0$:~\cite{jcp.128.114713}
\begin{equation}
\label{Eq:def_e_b_pol}
 \epsilon_\mathrm{b}^\mathrm{pol}=\sum_{i=1}^2 \epsilon_{\mathrm{b},i}^\mathrm{pol}=\sum_{i=1}^2\frac{g_i^2}{2|J|}\frac{1}{\sqrt{
\left(1+\frac{\hbar\omega_{\mathrm{p},i}}{2|J|}\right)^2-1}},
\end{equation}  
where $\epsilon_{\mathrm{b},i}^\mathrm{pol}$ are the contributions of high- and low-frequency phonon modes to the polaron binding energy.
The values of $g_1$ and $g_2$ in Table~\ref{Tab:model_params_gen} are obtained assuming that $\epsilon_\mathrm{b}^\mathrm{pol}=50$ meV and
$\epsilon_{\mathrm{b},1}^\mathrm{pol}=\epsilon_{\mathrm{b},2}^\mathrm{pol}$ and setting $|J|=125$ meV.

\subsection{Classification of Exciton States}
The classification of exciton states is unambiguous only for $J^\cb_{DA}=J^\vb_{DA}=0$
(noninteracting heterojunction),
when each exciton state $\psi^{x^{(0)}}_{(i\alpha_i)(j\beta_j)}$
can be classified as a donor exciton state (XD), a space-separated state,
an acceptor exciton state (XA) or a state in which the electron is in the donor, while the hole is in the acceptor (eDhA).
Because eDhA states are very well separated (in energy) from other groups of exciton states,
we will not further consider them.
In the group of space-separated states, CT and CS states can further
be distinguished by the mean electron-hole distance
\begin{equation}
 \langle r_{\mathrm{e-h}}\rangle_{x^{(0)}}=\sum_{\substack{i\alpha_i\\j\beta_j}}\left|\psi^{x^{(0)}}_{(i\alpha_i)(j\beta_j)}\right|^2 |i-j|.
\end{equation}
If the electron-hole interaction is set to zero, the mean electron-hole
distance for all the space-separated states is equal to $N$.
For the nonzero Coulomb interaction,
we consider a space-separated state as a CS state if its
mean electron-hole distance is larger than (or equal to) $N$, otherwise we
consider it as a CT state.

In general case, when at least one of $J^\cb_{DA},J^\vb_{DA}$ is different from zero (interacting heterojunction),
it is useful to explicitly separate the D/A interaction
from the interacting-carrier part of the
Hamiltonian (eq~\ref{Eq:H_c}),
\begin{equation}
\label{Eq:H_c_decom_large_LUMO_LUMO}
 H_\cb=H_\cb^{(0)}+H_{DA},
\end{equation}
where
\begin{equation}
\label{Eq:H_DA}
\begin{split}
 H_{DA}=&-J^\cb_{DA}\sum_{\substack{\beta_{N-1}\\\beta_N}}
\left(c_{(N-1)\beta_{N-1}}^\dagger c_{N\beta_N}+\mathrm{H.c.}\right)\\
&+J^\vb_{DA}\sum_{\substack{\alpha_{N-1}\\\alpha_N}}
\left(d_{(N-1)\alpha_{N-1}}^\dagger d_{N\alpha_N}+\mathrm{H.c.}\right),
\end{split}
\end{equation}
is the D/A interaction, and $H_\cb^{(0)}$ describes interacting carriers at the noninteracting
heterojunction. Exciton states of the noninteracting heterojunction
$\psi^{x^{(0)}}_{(i\alpha_i)(j\beta_j)}$ and corresponding exciton energies $\hbar\omega_{x^{(0)}}$
are obtained solving the electron-hole pair eigenproblem of $H_\cb^{(0)}$.
Exciton states of the interacting heterojunction $\psi^x_{(i\alpha_i)(j\beta_j)}$
are linear combinations of exciton states of the noninteracting heterojunction
\begin{equation}
\label{Eq:int_mix_nonint}
 \psi^x_{(i\alpha_i)(j\beta_j)}=\sum_{x^{(0)}}C_{xx^{(0)}}\psi^{x^{(0)}}_{(i\alpha_i)(j\beta_j)},
\end{equation}
and their character is obtained using this expansion. Namely, for each group $X^{(0)}$ of the exciton states of the
noninteracting heterojunction, we compute the overlap of state $x$ (of the interacting heterojunction)
with states belonging to this group
\begin{equation}
\label{Eq:amount_of_char}
 C_{X^{(0)}}^x=\sum_{x^{(0)}\in X^{(0)}}|C_{xx^{(0)}}|^2.
\end{equation}
The character of state $x$ is then the character of the group $X^{(0)}$ for which the overlap $C_{X^{(0)}}^x$ is maximum.

The electron in a space-separated state is predominantly located in the acceptor part of the heterojunction, while the hole is located in the donor part.
Since there is a number of single-electron levels per acceptor site, the electron in a space-separated state can be in different electronic bands originating from these
single-electron levels. A useful quantity for further classification of space-separated states is
\begin{equation}
 \label{Eq:p_beta}
 p_x(\beta)=\sum_{j=N}^{2N-1}\sum_{i\alpha_i}\left|\psi^x_{(i\alpha_i)(j\beta)}\right|^2,
\end{equation}
which represents the conditional probability that, given that the electron in state $x$ is in the acceptor, it belongs to the electronic band stemming from the single-electron level $\beta$.
The index of the electronic band $\beta_x$ to which the electron in space-separated state $x$ predominantly belongs is then the value of $\beta$ for which the
conditional probability $p_x$ is maximal.
In other words, space-separated state $x$ belongs to the CT$_{\beta_x}$ band.

Let us note here that, because of the large energy separation between the lower two (0 and 1) and the higher two (2 and 3) single-electron levels in the acceptor,
the electronic coupling $J^{\cb,\mathrm{ext}}_{A,12}$, which couples space-separated states belonging to CT$_0$ and CT$_1$ bands to the ones
belonging to CT$_2$ and CT$_3$ bands, is not effective. Therefore, the space-separated states from CT$_0$ and CT$_1$ bands are very weakly mixed with
(and essentially isolated from) space-separated
states of CT$_2$ and CT$_3$ bands, which permits us to separately analyze these two subgroups of space-separated states.

\subsection{Role of the Donor-Acceptor Coupling and the Resonant Mixing Mechanism}
In this section, we show that the D/A coupling is at the root of the resonant mixing mechanism which explains the presence of
space-separated (and XA) states that have a certain amount of donor character, can be reached by means of a photoexcitation,
and act as gateways to the space-separated manifold for the initial donor excitons.
However, the precise role of the D/A coupling is different in different energy regions of the exciton spectrum.
In the low-energy region of the exciton spectrum, which is dominated by the space-separated states belonging to CT$_0$ and CT$_1$ bands,
this coupling leads to the resonant mixing of two-particle (exciton) states.
On the other hand, in the high-energy region of the exciton spectrum, in which space-separated states belong to CT$_2$ and CT$_3$ bands,
it gives rise to the resonant mixing of single-electron states in the donor and acceptor.

To better appreciate the role of couplings $J^\cb_{DA},J^\vb_{DA}$, it is convenient to schematically represent exciton wave functions
$\psi^{x^{(0)}}_{(i\alpha_i)(j\beta_j)}$ and $\psi^{x}_{(i\alpha_i)(j\beta_j)}$ in the coordinate space. For the clarity of the discussion, we
assume that we have only one single-electron and single-hole state per site throughout the system.
This assumption does not compromise the validity of the conclusions to be presented in the case of more single-particle states
per site. On the abscissa of our coordinate
space is the hole coordinate, while the electron cordinate is on the ordinate.

\begin{figure}[htbp]
 \centering
 \includegraphics{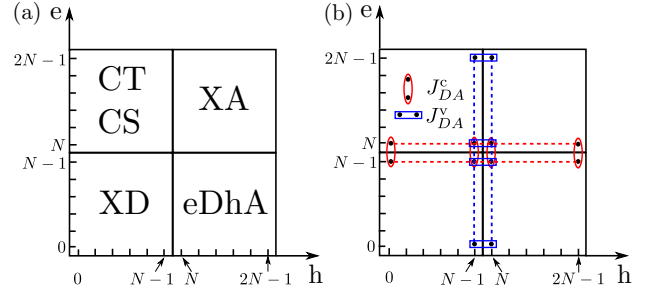}
 \caption{(a) At the noninteracting heterojunction, the wave function of each exciton state is confined to a single quadrant in the position space of the electron and hole.
(b) The points at which the sums in eq~\ref{Eq:h_barx0_x0} are evaluated:
the points relevant to the computation of the first and the second sum are grouped by red ellipses,
the points relevant to the computation of the third and the fourth sum are grouped by blue rectangles.}
 \label{Fig:model_pict}
\end{figure}

The wave functions of exciton states $x^{(0)}$ of the noninteracting heterojunction are confined to a single quadrant of our coordinate space, see Figure~\ref{Fig:model_pict}a.
For example, the wave function of a donor exciton state is nonzero only when both electron and hole coordinates are between 0 and $N-1$, and similarly for other
groups of exciton states. 
Because of the D/A interaction $H_{DA}$ (eq~\ref{Eq:H_DA}), exciton states $x$ of the interacting heterojunction
are mixtures of different exciton states $x^{(0)}$ of the noninteracting
heterojunction, see eq~\ref{Eq:int_mix_nonint}.
Therefore, the wave function of a general exciton state at the interacting heterojunction is not confined
to the quadrant which is in Figure~\ref{Fig:model_pict}a labeled by its prevalent character, but is nonzero also in other quadrants.
The D/A interaction $H_{DA}$ is written in the noninteracting-heterojunction exciton basis as
\begin{equation}
 H_{DA}=\sum_{\bar x^{(0)} x^{(0)}} h_{\bar x^{(0)} x^{(0)}} |\bar x^{(0)}\rangle\langle x^{(0)}|,
\end{equation}
with
\begin{equation}
\label{Eq:h_barx0_x0}
 \begin{split}
  h_{\bar x^{(0)} x^{(0)}}&=
-J^\cb_{DA}\sum_{\substack{k\alpha_k\\\beta_{N-1}\beta_N}}
  \psi^{\bar x^{(0)}*}_{(k\alpha_k)(N-1,\beta_{N-1})}\psi^{x^{(0)}}_{(k\alpha_k)(N\beta_N)}\\
&-J^\cb_{DA}\sum_{\substack{k\alpha_k\\\beta_{N-1}\beta_N}}
  \psi^{\bar x^{(0)}*}_{(k\alpha_k)(N\beta_N)}\psi^{x^{(0)}}_{(k\alpha_k)(N-1,\beta_{N-1})}\\
&+J^\vb_{DA}\sum_{\substack{k\beta_k\\\alpha_{N-1}\alpha_N}}
  \psi^{\bar x^{(0)}*}_{(N-1,\alpha_{N-1})(k\beta_k)}\psi^{x^{(0)}}_{(N\alpha_N)(k\beta_k)}\\
&+J^\vb_{DA}\sum_{\substack{k\beta_k\\\alpha_{N-1}\alpha_N}}
  \psi^{\bar x^{(0)}*}_{(N\alpha_N)(k\beta_k)}\psi^{x^{(0)}}_{(N-1,\alpha_{N-1})(k\beta_k)}.
 \end{split}
\end{equation}
The points at which the sums in the last equation (disregarding band indices) are to be evaluated are presented in Figure~\ref{Fig:model_pict}b.
The first two sums in eq~\ref{Eq:h_barx0_x0} are nonzero only when one state is of XD, and the other is of space-separated
character. Similarly, the other two sums in eq~\ref{Eq:h_barx0_x0} are nonzero only when one state is of XA,
and the other is of space-separated character.
Therefore, if $J^\cb_{DA}\neq 0$ and $J^\vb_{DA}=0$, XA states of the interacting heterojunction are identical
to XA states of the noninteracting heterojunction, while XD (space-separated) states of the interacting heterojunction
are generally combinations of XD and space-separated states of the noninteracting heterojunction.
Similarly, if $J^\cb_{DA}=0$ and $J^\vb_{DA}\neq 0$, XD states of the interacting heterojunction are identical
to XD states of the noninteracting heterojunction, while XA (space-separated) states of the interacting heterojunction
are generally combinations of XA and space-separated states of the noninteracting heterojunction.

The exact mechanism of this mixing is different in different parts of the exciton spectrum.
Let us start with the lower-energy part of the spectrum, which contains space-separated states belonging to CT$_0$ and CT$_1$ bands.
Single-electron states in the acceptor which originate from levels 0 and 1 do not exhibit strong resonant mixing with single-electron states in the donor,
thanks to the large energy separation between these two groups of states. Therefore, the relevant partitioning of the interacting-carrier Hamiltonian $H_\cb$
is the one embodied in eq~\ref{Eq:H_c_decom_large_LUMO_LUMO}.
Coefficients $C_{xx^{(0)}}$ in the expansion of exciton state $x$
(of the interacting heterojunction) in terms of exciton states $x^{(0)}$ (of the noninteracting heterojunction)
are obtained as solutions to the eigenvalue problem
\begin{equation}
\label{Eq:exc_eigen_prob_basis_ni_exc}
 \sum_{x^{(0)}}\left(\delta_{x^{(0)}\bar x^{(0)}}\hbar\omega_{x^{(0)}}
 +h_{\bar x^{(0)} x^{(0)}}\right) C_{xx^{(0)}}=\hbar\omega_x C_{x\bar x^{(0)}}.
\end{equation}
Since $h_{\bar x^{(0)} x^{(0)}}$ contains products of two exciton wave functions,
$|h_{\bar x^{(0)} x^{(0)}}|$ is generally much smaller than $|J^{\cb/\vb}_{DA}|$.
Therefore, most of the states in the lower-energy part of the interacting heterojunction are almost identical
to the respective states of the noninteracting heterojunction. However, whenever
$|h_{\bar x^{(0)} x^{(0)}}|\sim |\hbar\omega_{\bar x^{(0)}}-\hbar\omega_{x^{(0)}}|$,
there exists at least one state of the interacting heterojunction which is a mixture of states $\bar x^{(0)}$
and $x^{(0)}$ (which have different characters!) of the noninteracting heterojunction.
In other words, states $\bar x^{(0)}$ and $x^{(0)}$,
which are virtually resonant in energy,
exhibit resonant mixing to form the so-called bridge states of the interacting heterojunction.
Apart from their dominant character, which is obtained as previously explained,
bridge states also have nontrivial overlaps with noninteracting-heterojunction states of other characters.
For example, if $J^\vb_{DA}=0$, all the bridge states of the interacting heterojunction are
of mixed XD and space-separated character;
if $J^\cb_{DA}=0$, all the bridge states of the interacting heterojunction are
of mixed XA and space-separated character; if both couplings are nonzero,
bridge states of the interacting heterojunction are of mixed XD, XA, and space-separated character.
The emergence of bridge states in the low-energy part of the exciton spectrum requires subtle
energy alignment of exciton, i.e., two-particle, states. Bridge states formed by resonances between two-particle states are thus rather scarce.
Having a certain amount of the donor character, bridge states acquire oscillator strengths from donor states and can thus be directly generated from
the ground state. In the rest of our paper, it is convenient to consider as a bridge state any state (in the lower-energy part of the exciton spectrum)
of dominant CS, CT or XA character whose amount of donor character is at least 0.01.

On the other hand, in the high-energy region of the exciton spectrum, which contains space-separated states belonging to CT$_2$ and CT$_3$ bands,
there is significant mixing between single-electron states in the acceptor stemming from levels 2 and 3 and single-electron states in the donor.
In this case, instead of the decomposition of the interacting-carrier
part of the Hamiltonian given in eq~\ref{Eq:H_c_decom_large_LUMO_LUMO},
it is more convenient to separate the carrier-carrier interaction
(last three terms in eq~\ref{Eq:H_c}) from the part describing
noninteracting carriers (first two terms in eq~\ref{Eq:H_c}).
The latter part of the interacting-carrier Hamiltonian then gives rise to single-electron states of
the whole heterojunction which are delocalized on both the donor and acceptor as a consequence of the resonant mixing between single-electron states in the two materials.
Since one single-electron state of the entire system generally participates in many two-particle states,
exciton states having at least one carrier delocalized throughout the heterojunction are ubiquitous in the high-energy region of the spectrum.
They also generally have greater amount of donor character than the bridge states in the low-energy part of the spectrum, \emph{vide infra},
making them easily accessible from the ground state by a (suitable) photoexcitation.
The dominant character of these states can be different and to our further discussion
are relevant space-separated (CT and CS) states of CT$_2$ and CT$_3$ bands with partial donor character,
which will be further termed photon-absorbing charge-bridging (PACB) states.
This term has been repeatedly used in the literature to denote space-separated states
in which charges are delocalized throughout the system.~\cite{NanoLett.7.1967,PhysChemChemPhys.13.21461,PhysChemChemPhys.18.9514}
We note that the PACB states within our model do not have any other immediate relationship with PACB states reported in ab initio studies
of D/A interfaces apart from the charge-bridging property and relatively large oscillator strengths permitting their direct optical generation.

The bridge states owe their name to the fact that they indirectly connect,
via phonon-assisted processes,
a state of pure XD character to a state of pure space-separated character.
In our model, these two states cannot be involved in a single-phonon-assisted process
because of the form of exciton-phonon matrix elements
$\Gamma^{i\lambda_i}_{\bar x x}$ (eq~\ref{Eq:gamma}),
which contain products of exciton wave functions taken
at the same point.
Therefore, single-phonon-assisted transitons among exciton states
of the same character are most intensive and probable.
A state of pure XD character
can, however, also be coupled (via processes mediated by a donor phonon)
to a bridge state, which can then be coupled to a state of
pure space-separated character (via single-phonon processes mediated by acceptor phonons).
\begin{figure}[htbp]
 \centering
 \includegraphics{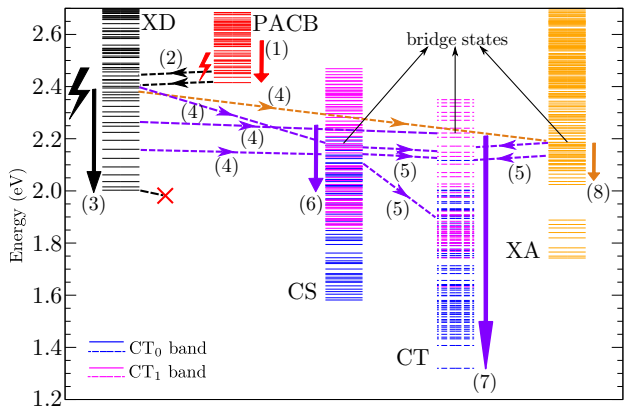}
 \caption{Exciton states relevant for our study divided in different groups.
In the third (the fourth) column (from the left), blue and magenta lines denote CS (CT) states belonging to CT$_0$ and CT$_1$ band, respectively.
Ultrafast exciton dynamics proceeds along the photophysical pathways denoted by (1)-(8), which are further specified in the Numerical Results section.
The solid arrows [pathways (1), (3), (6), (7), and (8)] indicate the deexcitation processes
occurring within one group of exciton states, whereas the dashed arrows [pathways (2), (4), and (5)] denote transitions
among different groups of exciton states. The black (red) bolt denotes the direct photoexcitation of excitons in donor (PACB) states.}
% (1) deexcitation within the PACB manifold;
% (2) transitions from the PACB to the XD manifold;
% (3) deexcitation within the XD manifold;
% (4) transitions from the XD manifold to bridge states;
% (5) transitions from bridge states to space-separated states;
% (6) not very pronounced deexcitation within the CS and XA manifolds;
% (7) deexcitation within the CT manifold towards low-energy CT states.

 \label{Fig:menagerie}
\end{figure}

In the remaining part of our study, we will for ease of presentation adopt the following classification of the exciton states.
Since space-separated states belonging to CT$_2$ and CT$_3$ bands which are relevant to our study are PACB states,
we will not discriminate between CT and CS states in CT$_2$ and CT$_3$ bands, but rather refer to all of them as PACB states. 
We will, however, distinguish between CS and CT states in CT$_0$ and CT$_1$ bands and, for brevity of discussion,
we will denote them simply as CS and CT states. This classification facilitates the understanding of
the role that PACB states play in ultrafast interfacial dynamics by enabling direct comparison between results obtained with
all four and only two lower orbitals per acceptor site, {\it vide infra}.
The comparison is plausible since there is a well defined correspondence between XA, CT, and CS states in the lower-energy part of the exciton spectrum
(four orbitals per acceptor site) and the corresponding states when only two orbitals per acceptor site are taken into account.
The part of the exciton spectrum which is relevant for our study is shown in Figure~\ref{Fig:menagerie}. 

\section{Numerical Results}
\label{Sec:num_res}
In this section, we present results for the exciton dynamics at the model heterojunction
during and after its pulsed excitation. The form of the excitation is
\begin{equation}
 E(t)=E_0\cos(\omega_c t)\exp\left(-\frac{t^2}{\tau_G^2}\right)
\theta(t+t_0)\theta(t_0-t),
\end{equation}
where $\omega_c$ is its central frequency, $2t_0$ is its duration,
$\tau_G$ is the characteristic time of the Gaussian envelope,
and $\theta(t)$ is the Heaviside step function.
In all the computations, we set $t_0=50$ fs and $\tau_G=20$ fs.
Computing the energy- and time-resolved exciton populations $\varphi_X(E,t)$ (eqs~\ref{Eq:ene_res_coh} and~\ref{Eq:ene_res_incoh})
or the exciton DOS, we represent $\delta$ functions by a Gaussian
with the standard deviation of 10 meV. 

We start with the analysis of the ultrafast exciton dynamics
when model parameters assume the values listed in Table~\ref{Tab:model_params_gen}
and the system is excited at the bright donor state located around
$\hbar\omega_c=2.35$ eV, which is significantly above the lowest donor state.
We also present the results obtained taking into account only two lower single-electron
levels (of energies $\epsilon^\cb_{A,0}$ and $\epsilon^\cb_{A,1}$) in the acceptor per site,
while the values of all other model parameters are as listed in Table~\ref{Tab:model_params_gen}.
The comparison of these results helps us understand the effects that the presence of two higher single-electron levels
in the acceptor has on ultrafast exciton dynamics in our model.

\begin{figure}[htbp]
 \centering
 \includegraphics{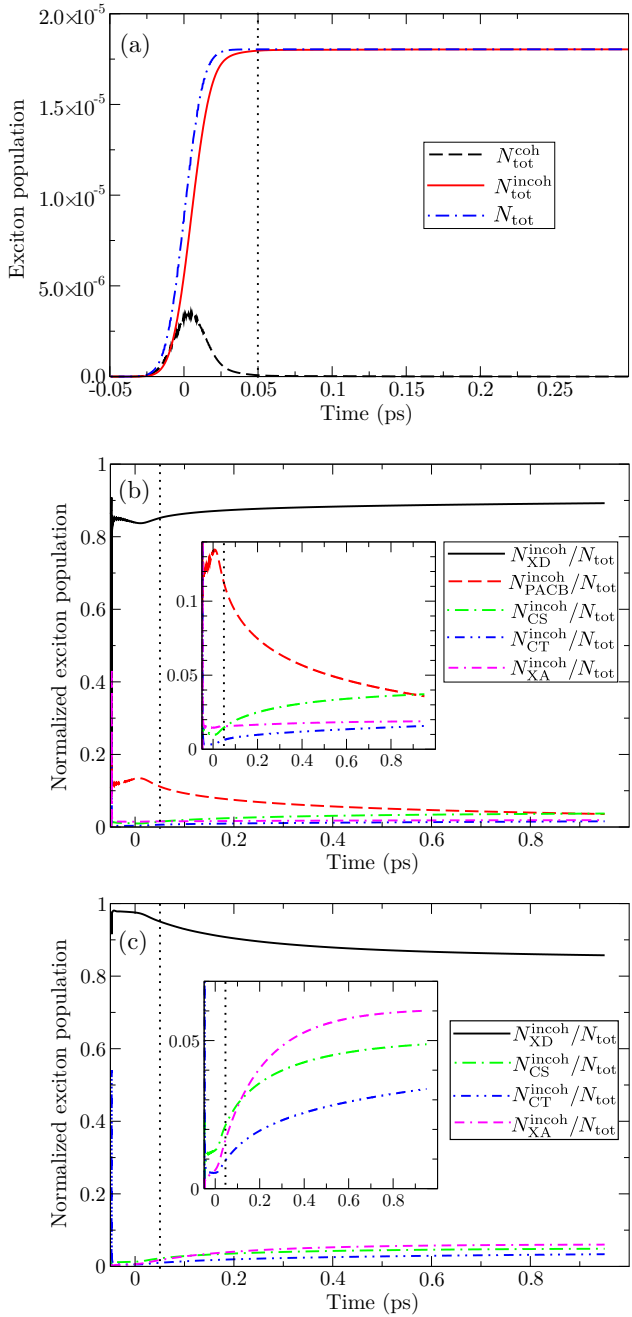}
 \caption{Time dependence of (a) the total exciton population and its coherent and incoherent parts,
(b,c) normalized incoherent populations of different groups of exciton states.
In panels a and b, we take four single-electron levels per acceptor site, while in panel c we take only two lower
single-electron levels ($\epsilon^\cb_{A,0}$ and $\epsilon^\cb_{A,1}$) per acceptor site.
The dotted vertical lines denote the end of the excitation.}
 \label{Fig:coh_incoh_tot_incoh_xd_cs_ct_xa}
\end{figure}

\begin{figure}[htbp]
 \centering
 \includegraphics{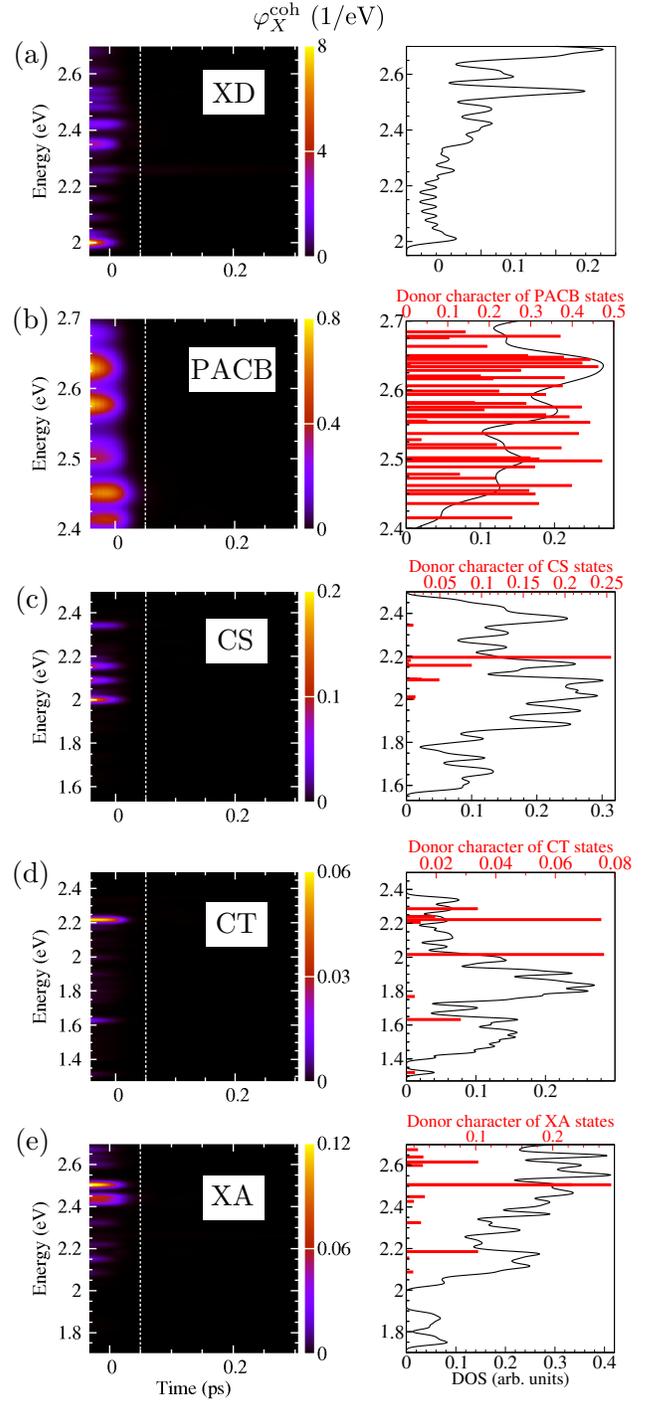}
 \caption{Density plots of $\varphi^\mathrm{coh}_X(E,t)$ for
(a) XD, (b) PACB, (c) CS,
(d) CT, and (e) XA states.
Each density plot is complemented with the plot of the corresponding exciton DOS. In b-e,
exciton DOS plots contain amounts of the donor character of exciton states
(see eq~\ref{Eq:amount_of_char}) [in panels c-e, as long as it is greater than 0.01].
}
 \label{Fig:ene_res_coh}
\end{figure}

In Figure~\ref{Fig:coh_incoh_tot_incoh_xd_cs_ct_xa}a we show the time dependence of the total coherent exciton population
$N_\mathrm{tot}^\mathrm{coh}=\sum_x |y_x|^2$,
total incoherent exciton population $N_\mathrm{tot}^\mathrm{incoh}=\sum_x \bar n_{xx}$, and total exciton
population $N_\mathrm{tot}$ (eq~\ref{Eq:pop_total}).
Exciting well above the lowest donor state,
the conversion from coherent to incoherent exciton populations is rapid
and is completed in a couple of tens of femtoseconds after the end of the pulsed excitation.
Figure~\ref{Fig:ene_res_coh}a-e presents density plots of energy- and time-resolved distributions
$\varphi^\mathrm{coh}_X(E,t)$ of coherent exciton populations for different groups of exciton states $X$.
Comparing the ranges of color bars in Figure~\ref{Fig:ene_res_coh}a-e,
we conclude that the excitation predominantly generates donor excitons.
We observe in Figure~\ref{Fig:ene_res_coh}a that the initially populated donor states are
the states located around 2.35 and 2.42 eV, together with the lowest donor state at around 2 eV.
Even though we pump well above the lowest donor state, this state is prone to the direct optical generation
because of its very large dipole moment $M_x$ (eq~\ref{Eq:M_x})
for direct generation from the ground state and the spectral width of the pulse. 
Apart from donor states, PACB states are also initially populated, see Figure~\ref{Fig:ene_res_coh}b.
In Figure~\ref{Fig:ene_res_coh}c-e we see that energy positions of the bright spots in the density plots on the left
correspond very well to the energy positions of red bars, which indicate bridge states of dominant CS, CT, and XA character, on the right.
In other words, these states can be directly optically generated from the ground state,
as already discussed.

The time dependence of normalized incoherent populations of different groups of exciton states
is presented in Figure~\ref{Fig:coh_incoh_tot_incoh_xd_cs_ct_xa}b.
Figure~\ref{Fig:coh_incoh_tot_incoh_xd_cs_ct_xa}c shows normalized incoherent populations in the model with only
two accessible electronic states (of energies $\epsilon^\cb_{A,0}$ and $\epsilon^\cb_{A,1}$) at each acceptor site.
Comparing panels b and c of Figure~\ref{Fig:coh_incoh_tot_incoh_xd_cs_ct_xa}, we conclude that
the presence of PACB states significantly affects exciton dynamics on ultrafast time scales.
In the presence of only two lower electronic levels in the acceptor,  
the number of donor excitons decreases, while
the numbers of CS, CT, and XA excitons increase after the excitation (see Figure~\ref{Fig:coh_incoh_tot_incoh_xd_cs_ct_xa}c).
On the other hand, taking into account the presence of higher-lying electronic orbitals in the acceptor and pumping well above the lowest donor exciton,
the populations of XD, XA, CT and CS states
increase, while the population of PACB states decreases after the excitation.
The fact that donor states acquire population after the end of the pulse may at first seem counterintuitive, since initially generated donor excitons are expected
to dissociate, performing transitions to the space-separated manifold.
Having significant amount of donor character, PACB states
are well coupled (via single-phonon-assisted processes) to the manifold of donor excitons, while their coupling to space-separated states belonging to
CT$_0$ and CT$_1$ bands is essentially negligible (see also the paragraph following eq~\ref{Eq:p_beta}).
Therefore, instead of performing single-phonon-assisted transitions to lower-energy space-separated states, initially generated PACB excitons
perform transitions toward donor states, i.e., the number of donor excitons increases at the expense of excitons initially generated in PACB states.
While, at the end of the pulse, excitons in PACB states comprise around 11\% of the total exciton population, 900 fs after the pulse their participation
in the total population reduces to 4\%. At the same time, the normalized number of donor excitons increases from around 85\% to around 89\% of
the total exciton population, meaning that some of the donor excitons are converted into XA, CT, and CS states,
which is seen in Figure~\ref{Fig:coh_incoh_tot_incoh_xd_cs_ct_xa}b as the increase in the populations of these states.

In the model with four accessible electronic orbitals per acceptor site,
the major part of space-separated states that are populated on 100-fs time scales following the excitation are directly generated PACB states.
This conclusion is in line with our recent results regarding ultrafast photophysics in a model where the LUMO-LUMO offset is comparable to
the effective bandwidth of the LUMO band of the acceptor.~\cite{PhysRevB.95.075308}
Namely, we have recognized that the resonant mixing between single-electron states in the LUMO bands of the two materials
is at the root of the ultrafast direct optical generation of space-separated charges.
Here, the same mechanism is responsible for the observed direct generation of excitons in PACB states, which now acquire nonzero oscillator strengths
due to the energy alignment between single-electron states stemming from the donor LUMO orbital and higher-than-LUMO acceptor orbitals.
On the contrary, if only electronic orbitals close to the LUMO orbital are taken into account,
populations of space-separated states present on 100-fs time scales after the excitation mainly reside in bridge states, which are formed
by two-particle resonant mixing.
The populations of bridge states are dominantly built by phonon-assisted transitions
from initially generated donor excitons (since the direct generation of excitons in bridge states is not very pronounced for the excitation studied).
Therefore, in our model, the PACB states can enhance the generation of space-separated charges on ultrafast time scales
by allowing for their direct optical generation and not by acting as intermediate states of charge separation starting from initial donor excitons.

\begin{figure}[h!]
 \centering
 \includegraphics{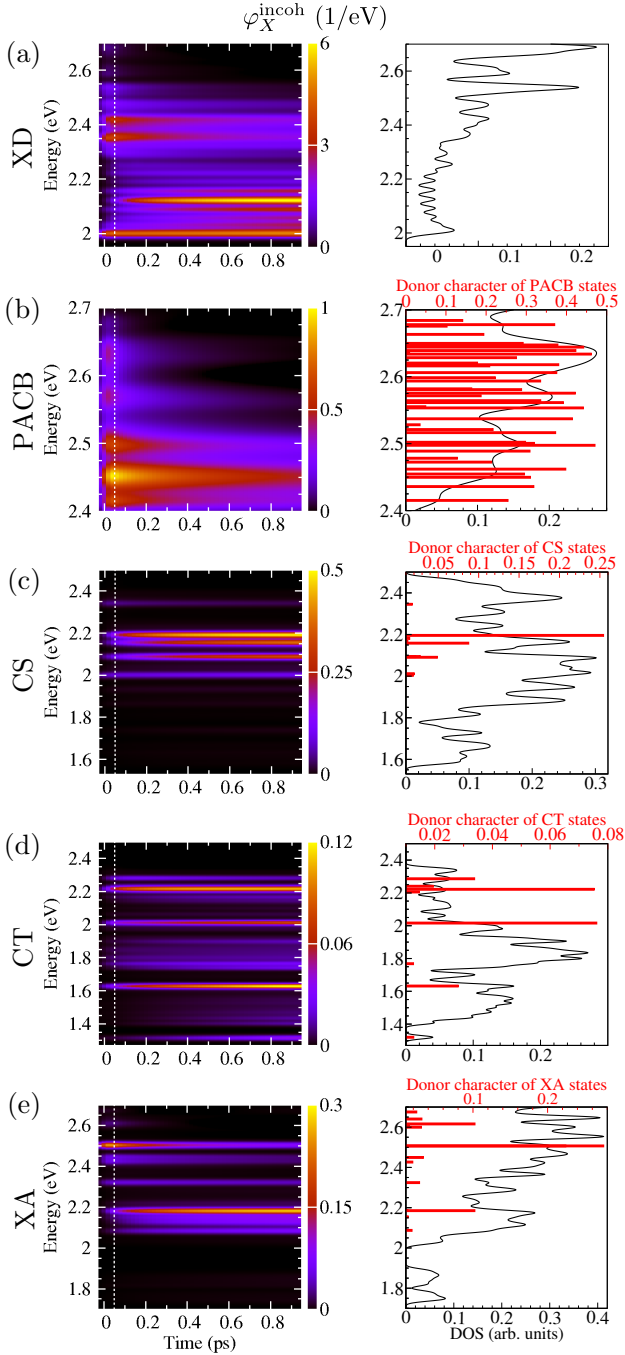}
 \caption{Density plots of $\varphi_X^\mathrm{incoh}(E,t)$ for
(a) XD, (b) PACB, (c) CS,
(d) CT, and (e) XA states.
Each density plot is accompanied by the plot of the corresponding exciton DOS. In b-e,
the exciton DOS plots contain the amount of the donor character of exciton states
(see eq~\ref{Eq:amount_of_char}) [in panels c-e, as long as it is greater than 0.01].}
 \label{Fig:ene_res}
\end{figure}

In order to understand the photophysical pathways of ultrafast exciton dynamics, in Figure~\ref{Fig:ene_res}a-e
we depict the density plots of $\varphi_X^\mathrm{incoh}(E,t)$ for various groups $X$ of exciton states.
For the completeness of the discussion, in the Supporting Information we provide the density plots of $\varphi_X^\mathrm{incoh}(E,t)$
in the model with only two lower electronic orbitals per acceptor site and compare them to the plots presented here.
As already explained,
the excitons initially generated in PACB states (red bolt in Figure~\ref{Fig:menagerie}) undergo deexcitation
within the PACB manifold (pathway (1) in Figure~\ref{Fig:menagerie})
followed by phonon-mediated transitions toward the manifold of donor states (pathway (2) in Figure~\ref{Fig:menagerie}; see Figure~\ref{Fig:ene_res}b).
Donor excitons (either the ones initially generated in higher-lying bright states (black bolt in Figure~\ref{Fig:menagerie}) or the ones originating from PACB excitons)
are involved in a series of ultrafast phonon-assisted transitions toward lower-energy states.
Most of these transitions happen within the XD manifold (pathway (3) in Figure~\ref{Fig:menagerie};
see the series of more or less bright bands in the density plot of Figure~\ref{Fig:ene_res}a),
which is consistent with the fact that donor excitons comprise the largest part of the total
exciton population at every instant.
The deexcitation within the XD manifold proceeds 
until the lowest XD state is reached. In fact,
we see that already for $t\gtrsim 250$ fs, XD population resides mainly in the lowest donor state
at around 2 eV and the donor state at around 2.13 eV. 
The lowest donor state is almost uncoupled from the space-separated manifold,
acting as a trap state for exciton dissociation,
which is in line with other studies.~\cite{JPhysChemLett.6.1702}
The other donor state (at around 2.13 eV)
acting as a trap state for exciton dissociation is specific to our computation.

\begin{figure}[h!]
 \centering
 \includegraphics{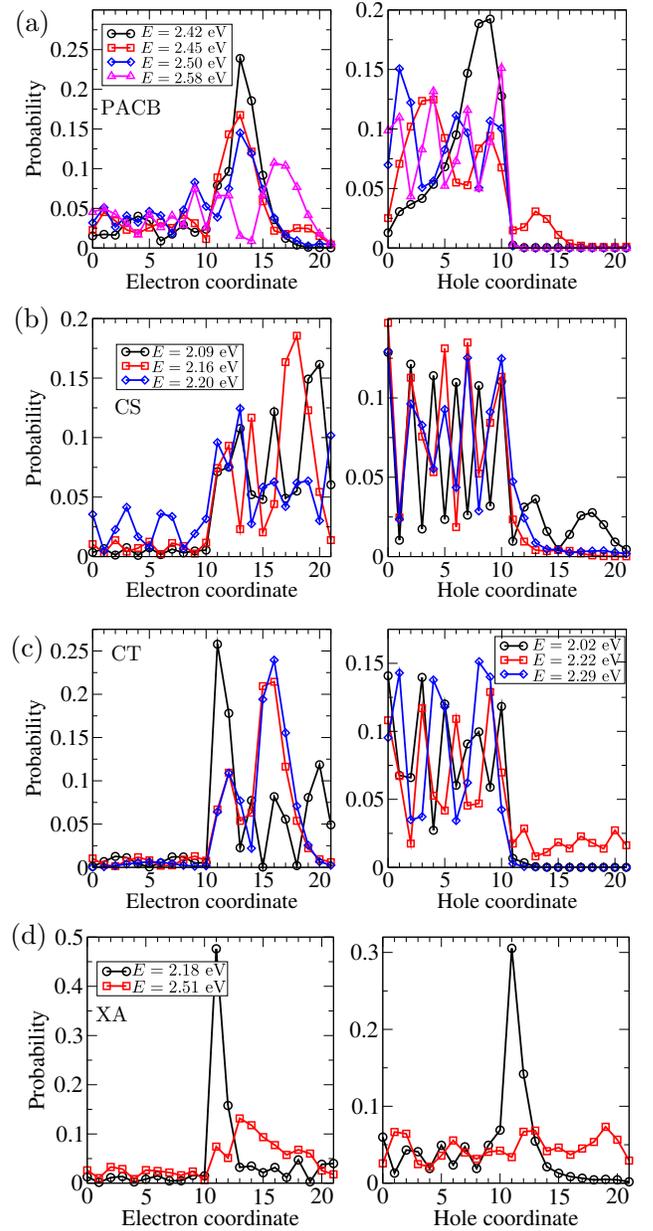}
 \caption{Probability distributions of the electron (left) and hole (right)
in representative (a) PACB states and bridge states of dominant (b) CS,
(c) CT, and (d) XA character.}
 \label{Fig:bridge_delocalization}
\end{figure}

In the course of the deexcitation from
the higher-lying donor states and before reaching a trap state for exciton dissociation,
a donor exciton can perform a transition to a bridge state (pathway (4) in Figure~\ref{Fig:menagerie}).
As seen in Figure~\ref{Fig:ene_res}c-e, the energy positions of the
bright bands in the density plots on the left
match exactly the energy positions of red bars displaying the amount of donor character
of dominantly space-separated or XA states on the right.
Figure~\ref{Fig:bridge_delocalization}b-d
depicts probability distributions of the electron and hole in representative bridge states of different
dominant characters, while Figure~\ref{Fig:bridge_delocalization}a shows the same quantities for particular PACB states.
All the bridge states exhibit carrier delocalization throughout
the system; this property makes them accessible from the initial states of donor excitons.
The same holds for PACB states: since the carriers in these states are delocalized throughout the heterojunction, these states
inherit oscillator strengths from donor excitons and may thus be directly accessed by an optical excitation.
Moreover, this property enables efficient phonon-assisted coupling between PACB states and donor states.
The bridge states gain significant populations during the first 100 fs following the excitation (pathway (4) in Figure~\ref{Fig:menagerie})
and concomitantly the excitons initially generated in PACB states
perform phonon-mediated transitons toward donor states (pathway (2) in Figure~\ref{Fig:menagerie}).

Once the exciton has reached a bridge state,
it can deexcite within the manifold of its dominant character (pathways (6)-(8) in Figure~\ref{Fig:menagerie})
or it can perform a transiton to the CT manifold (pathway (5) in Figure~\ref{Fig:menagerie})
followed by a number of downward transitions within this manifold (pathway (7) in Figure~\ref{Fig:menagerie}; see the series of more or less bright bands
between 1.3 and 2.2 eV in the density plot of Figure~\ref{Fig:ene_res}d).
The gradual deexcitation within the CT manifold
leads to the delayed buildup of populations of
low-energy CT states (pathway (7) in Figure~\ref{Fig:menagerie}; see bright bands at around 1.62 and 1.32 eV
in the density plot in Figure~\ref{Fig:ene_res}d), which happens on a picosecond time scale.
Apart from mediating the charge separation, bridge states can also act as competing final states.
In our computation, at every instant, virtually all CS excitons reside in bridge states of dominant
CS character, and the progressive deexcitation within the CS manifold (pathway (6) in Figure~\ref{Fig:menagerie}) is not pronounced
(see Figure~\ref{Fig:ene_res}c).
Analogous situation is observed analyzing the energy- and time-resolved populations of
XA states (pathway (8) in Figure~\ref{Fig:menagerie}) in Figure~\ref{Fig:ene_res}e.
This 2-fold role of bridge states observed in
our computations is in agreement with conclusions of previous studies.~\cite{PhysRevLett.100.107402}

\subsection{Ultrafast Exciton Dynamics for Various Central Frequencies}
\begin{figure}[h!]
 \centering
 \includegraphics{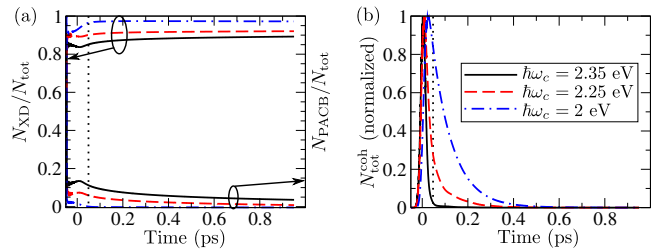}
 \caption{Time dependence of (a) the normalized number of excitons in donor and PACB states, and
 (b) the total coherent exciton population,
 for different central frequencies of the excitation.
For convenience, the total coherent population shown in panel b is normalized so that its maximal value is equal to 1.}
 \label{Fig:diff_omegac}
\end{figure}

The exact photophysical pathways along which the exciton dynamics proceeds on ultrafast time scales
strongly depend on the frequency of the excitation, the exciton dissociation being more pronounced for
larger excess energy.~\cite{nmat12-29,JPhysChemC.118.28527} 
Here, we examine ultrafast exciton dynamics for three different excitations of central frequencies
$\hbar\omega_c=2.35$, 2.25, and 2 eV (excitation at the lowest donor state).
As the central frequency of the excitation is decreased, i.e.,
as the initially generated donor excitons are closer in energy to the lowest donor state,
the conversion from coherent to incoherent exciton population is slower and the time scale
on which exciton coherences with the ground state dominate the interfacial dynamics is longer (see Figure~\ref{Fig:diff_omegac}b).
At the same time, the participation of excitons in PACB states in the total exciton population is decreased,
whereas donor excitons comprise larger part of the total population (see Figure~\ref{Fig:diff_omegac}a).
Namely, as the central frequency is lowered toward the lowest donor state, the initial optical generation of excitons
in PACB states is less pronounced and the pathways (1) and (2) in Figure~\ref{Fig:menagerie}
become less important, while the possible photophysical pathways of the initially generated donor excitons
become less diverse. Therefore, the phonon-assisted processes responsible for the conversion from coherent to incoherent exciton
populations and for the ultrafast phonon-mediated transitions from donor states toward space-separated states are less effective.
As a consequence, the conversion from coherent to incoherent exciton populations is slower, and initially generated donor excitons
tend to remain within the manifold of donor states (pathway (3) in Figure~\ref{Fig:menagerie}, down to the lowest donor state, is preferred to
pathways (4)-(7), which may lead to space-separated states). The latter fact is especially pronounced exciting at the lowest donor state,
which is very weakly coupled to the space-separated manifold, when around 80\% of the total exciton population lies in the lowest donor state,
meaning that the ultrafast charge transfer upon excitation at this state is not significant.
In the Supporting Information, we present the density plots
of $\varphi^\mathrm{incoh}_X(E,t)$ for different groups of exciton states $X$ and excitations of different central frequencies.

\subsection{Influence of Small Variations of the LUMO-LUMO Offset on Ultrafast Exciton Dynamics}
In this work, we deal with rather large LUMO-LUMO offsets, when the bridge states
emerge as a consequence of the energy resonance between two-particle (exciton) states.
The energies of these states, as well as their number and amount of the donor character,
are therefore very sensitive to the particular exciton energy level alignment at the heterojunction.
On the other hand, the properties of PACB states are not expected to be particularly sensitive to the details of the energy level alignment, since they originate
from resonances between single-electron states in the donor and acceptor.
In order to demonstrate this difference between bridge states
and PACB states, we performed computations with different, but very close, values of the LUMO-LUMO offset.
The LUMO-LUMO offset is varied by changing all the parameters
$\epsilon^\cb_{A,0},\epsilon^\cb_{A,1},\epsilon^\cb_{A,2},\epsilon^\cb_{A,3},\epsilon^\vb_{A,0}$ in Table~\ref{Tab:model_params_gen}
by the same amount, keeping all the other model parameters fixed.
The effects of small variations of LUMO-LUMO offset are studied for $J^\vb_{DA}=0$, when all the bridge states are of mixed XD and space-separated character
and, since $d^{\cb\vb}_A=0$, XA states do not participate in the ultrafast exciton dynamics.
The exclusion of XA states from the dynamics significantly decreases the numerical effort and at the same time
allows us to concentrate on the dynamics of ultrafast electron transfer, instead of considering both electron transfer
and exciton transfer. The main qualitative features of the ultrafast exciton dynamics described earlier
remain the same, as detailed in the Supporting Information.
\begin{figure}[h!]
 \centering
 \includegraphics{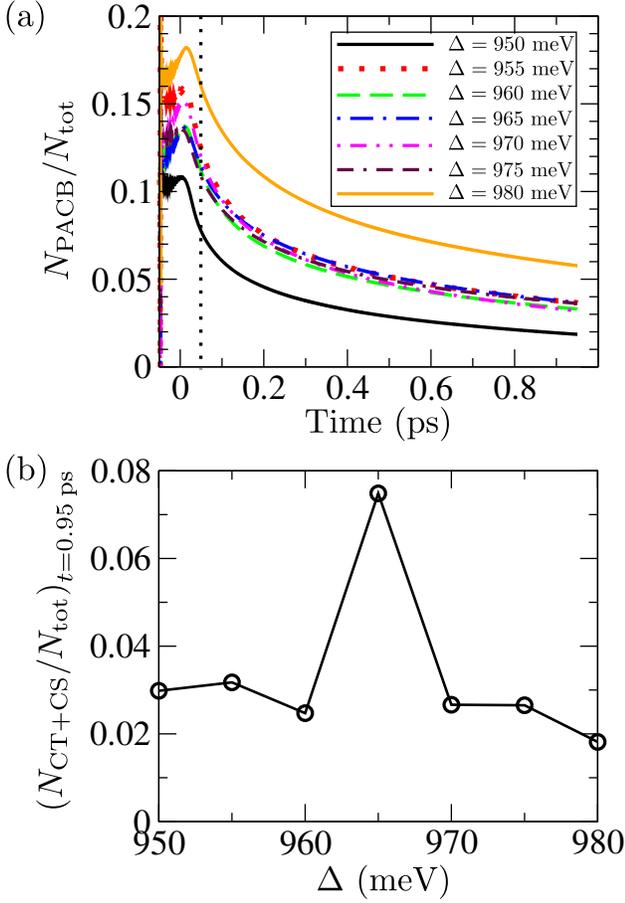}
 \caption{(a) Time dependence of the normalized number of PACB excitons for different LUMO-LUMO offsets $\Delta$.
(b) The relative number of excitons in space-separated (CT and CS) states
900 fs after the excitation for different LUMO-LUMO offsets $\Delta$.}
 \label{Fig:diff_lumo_lumo}
\end{figure}
The system is excited at $\hbar\omega_c=2.35$ eV.
Figure~\ref{Fig:diff_lumo_lumo}a presents the time dependence of the normalized number excitons in PACB states,
while Figure~\ref{Fig:diff_lumo_lumo}b shows the normalized number of excitons in space-separated states
900 fs after the excitation for different
LUMO-LUMO offsets ranging from 950 to 980 meV in steps of 5 meV.
Small variations of the LUMO-LUMO offset between 955 and 975 meV weakly affect the portion of PACB excitons in the total exciton population.
However, for the LUMO-LUMO offset of 980 meV, the normalized number of excitons in PACB states is somewhat higher than for the other considered values,
while this number is somewhat smaller for the LUMO-LUMO offset of 950 meV.
Namely, for larger LUMO-LUMO offsets, the lowest state of CT$_2$ band is closer to the central frequency of the excitation,
and the direct optical generation of excitons in PACB states is more pronounced.
For smaller LUMO-LUMO offsets, the initial generation of excitons in PACB states is to a certain extent suppressed because the energy difference between the lowest
state of CT$_2$ band and the central frequency of the excitation is larger.
On the other hand, the relative number of
space-separated excitons
can change up to three times as a result of small changes in the LUMO-LUMO offset.
The different behavior displayed by the relative numbers of PACB excitons and space-separated excitons
is a consequence of different mechanisms by which PACB states and bridge states emerge.
The peak in the normalized number of space-separated excitons observed for the LUMO-LUMO offset
of 965 meV signalizes that the exciton-level alignment at this point favors either (i) formation of more bridge states of dominant
space-separated character
than at other points or
(ii) formation of bridge states that couple more strongly to initial donor states than bridge states at other points.  

\subsection{Influence of Carrier-Phonon Interaction Strength and Temperature on Ultrafast Exciton Dynamics}
We have analyzed the ultrafast exciton dynamics for different strengths of the carrier-phonon coupling,
exciting the system at $\hbar\omega_c=2.35$ eV. The polaron binding energy $\epsilon^\mathrm{b}_\mathrm{pol}$ (eq~\ref{Eq:def_e_b_pol}),
which is a measure of the carrier-phonon interaction strength,
assumes values of 20, 50, and 70 meV.
\begin{figure}[h!]
 \centering
 \includegraphics{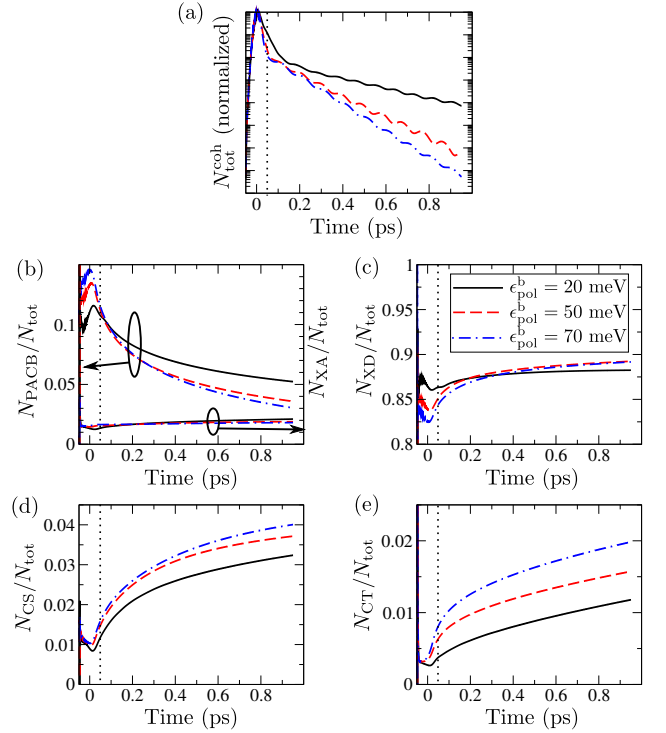}
 \caption{(a) Time dependence of the total coherent exciton population $N^\mathrm{coh}_\mathrm{tot}$
for different carrier-phonon interaction strengths. For convenience, $N^\mathrm{coh}_\mathrm{tot}$ is normalized
so that its maximum assumes the same value for all studied interaction strengths.
Dynamics of normalized incoherent excition populations of
(b) PACB and XA, (c) XD, (d) CS, and (e) CT states,
for different interaction strengths.}
 \label{Fig:e_ph_infl_1}
\end{figure}
Since the carrier-phonon interaction mediates the conversion from coherent to incoherent exciton populations,
weaker carrier-phonon coupling makes this conversion somewhat slower (see Figure~\ref{Fig:e_ph_infl_1}a).
We note that, for all the interaction strengths considered,
the total coherent population decays 100 times (compared to its maximal value) in $\lesssim 100$ fs following the excitation, meaning that
the conversion is in all three cases relatively fast.

The normalized number of excitons in PACB states is smaller for stronger carrier-phonon interaction (see Figure~\ref{Fig:e_ph_infl_1}b).
The characteristic time scale for the decay of the population of PACB states is shorter for stronger carrier-phonon interaction,
which is a consequence of stronger phonon-mediated coupling among PACB states and donor states (pathway (2) in Figure~\ref{Fig:menagerie}).
For larger interaction strength, the populations of CS and CT states
comprise larger part of the total exciton population (see Figure~\ref{Fig:e_ph_infl_1}d,e).
Namely, the stronger is the carrier-phonon interaction,
the more probable are the transitions from donor excitons to bridge states (pathway (4) in Figure~\ref{Fig:menagerie}) and the larger are the populations
of CS and CT states (pathways (5)-(7) in Figure~\ref{Fig:menagerie}).
The relative number of acceptor excitons does not change very much with the carrier-phonon interaction strength (see Figure~\ref{Fig:e_ph_infl_1}b).
The variation in the relative number of donor excitons brought about by the changes in the interaction strength is governed by a number of competing factors.
First, stronger carrier-phonon interaction favors larger number of donor excitons, since phonon-assisted transitions from
PACB to XD states (pathway (2) in Figure~\ref{Fig:menagerie}) are more pronounced.
Second, for stronger interaction, the transitions from XD to bridge states
are more probable (pathway (4) in Figure~\ref{Fig:menagerie}).
Third, since phonon-mediated transitions are most pronounced between exciton states of the same character, stronger interaction may also favor
deexcitation of donor populations within the XD manifold (down to the lowest XD state, pathway (3) in Figure~\ref{Fig:menagerie})
to possible transitions (via bridge states) to the space-separated manifold
(pathways (4)-(7) in Figure~\ref{Fig:menagerie}).
From Figure~\ref{Fig:e_ph_infl_1}c we see that, as a result of all these factors, the relative number of donor excitons does not change monotonously with the interaction strength.

\begin{figure*}
 \centering
 \includegraphics{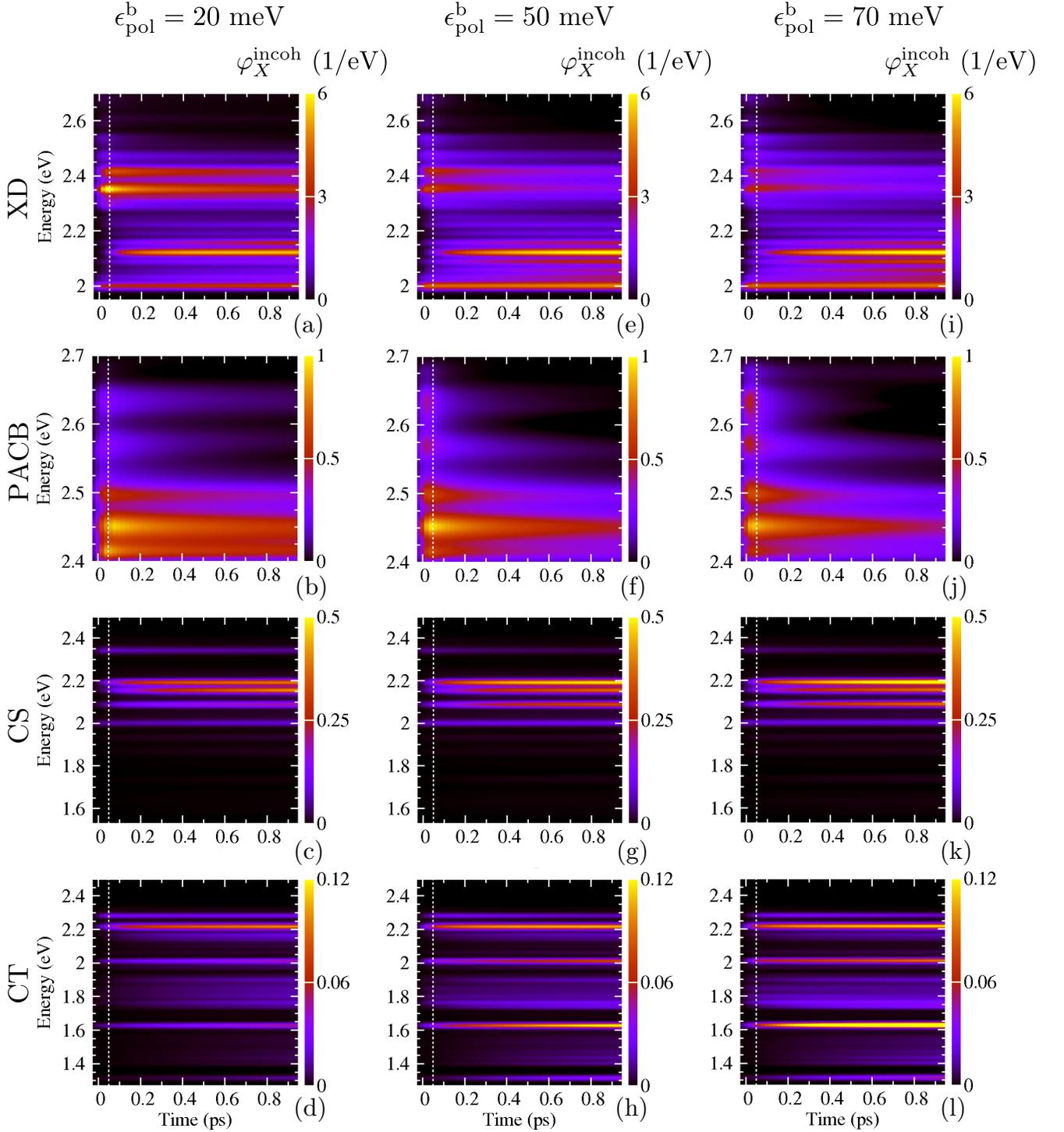}
 \caption{Energy- and time-resolved incoherent exciton populations $\varphi^\mathrm{incoh}_X(E,t)$
for different carrier-phonon interaction strengths: (a,b,c,d): $g_1=26.7$ meV, $g_2=59.7$ meV;
(e,f,g,h): $g_1=42.2$ meV, $g_2=94.3$ meV; (i,j,k,l): $g_1=54.0$ meV, $g_2=111.6$ meV.
Groups of exciton states: (a,e,i): XD states; (b,f,j): PACB states;
(c,g,k): CS states;
(d,h,l): CT states.}
 \label{Fig:e_ph_details}
\end{figure*}

In order to understand how the changes in carrier-phonon interaction strength affect the photophysical pathways along which the ultrafast exciton dynamics proceeds,
in Figure~\ref{Fig:e_ph_details}a-l we present energy- and time-resolved incoherent populations of various groups of exciton states (in different rows)
and for different interaction strengths (in different columns).
While for the strongest studied interaction initially generated higher-lying donor excitons and excitons in PACB states
leave the initial states rapidly (see Figure~\ref{Fig:e_ph_details}i,j), for the weakest studied interaction strength
significant exciton population remains in these states
during the first picosecond of the exciton dynamics (see Figure~\ref{Fig:e_ph_details}a,b).
The deexcitation of donor excitons takes place predominantly
within the XD manifold (pathway (3) in Figure~\ref{Fig:menagerie})
for all three interaction strengths, compare the ranges of color bars in Figures~\ref{Fig:e_ph_details}a,e,i.
For the weakest studied interaction, the lowest donor state, which is a trap for the exciton dissociation, is largely bypassed in the
course of the deexcitation, whereas for stronger carrier-phonon interactions this state acquires significant population already from the beginning of the excitation.
Energy- and time-resolved populations of CS states
are very nearly the same for all three interaction strengths studied
(see Figures~\ref{Fig:e_ph_details}c,g,k).
The major part of the CS population resides in bridge states, and the deexcitation within the subset of CS states (pathway (6) in Figure~\ref{Fig:menagerie}) is not
very pronounced. On the other hand, the deexcitation within the subset of CT states (pathway (7) in Figure~\ref{Fig:menagerie}),
down to the lowest CT state,
is observed for all the interaction strengths considered (see the series of more or less bright bands in Figure~\ref{Fig:e_ph_details}d,h,l).
While for the weakest interaction the largest portion of the CT population resides in the bridge state of CT character located at around 2.2 eV,
for the strongest interaction the major part of the CT population is located in the lowest state of CT$_1$ band at around 1.63 eV.

The carrier-phonon coupling thus acts in two different ways. On the one hand, stronger carrier-phonon interaction enhances exciton dissociation
and subsequent charge separation by (i) enabling phonon-assisted transitons from a donor state to space-separated
states via bridge states (pathways (4) and (5) in Figure~\ref{Fig:menagerie}) and (ii) enabling phonon-assisted transitions within the space-separated manifold once
a space-separated state is reached (pathways (6) and (7) in Figure~\ref{Fig:menagerie}).
On the other hand, stronger carrier-phonon coupling is detrimental to exciton dissociation and further charge separation
because (i) it makes donor states more easily accessible from initially generated PACB excitons (pathway (2) in Figure~\ref{Fig:menagerie}), and,
similarly, it may favor backward transitions from a bridge state to a donor state with respect to transitions to the space-separated manifold
and (ii) downward phonon-assisted transitions make low-energy CT states,
which are usually considered as traps for charge separation,
populated on a picosecond time scale following the excitation (pathway (7) in Figure~\ref{Fig:menagerie}).

In the Supporting Information we examine the temperature dependence of the ultrafast heterojunction dynamics.
We find that the effect of temperature variations on exciton dynamics occurring on subpicosecond time scales
is not particularly pronounced, as has been repeatedly recognized in the literature.~\cite{JPhysChemLett.1.2255,JChemPhys.140.044104,ChemPhys.442.111}

\section{Discussion and Conclusion}
Using a relatively simple, but physically grounded model of an all-organic heterointerface,
we have investigated subpicosecond dynamics of exciton dissociation and charge separation in the framework of the density matrix theory complemented with the DCT scheme.
Our model is constructed as an effective model intended to describe the dynamics of excitation transport in the direction perpendicular to the interface
and it is parametrized using the literature data for the P3HT/PCBM blend.

Apart from the electronic states of the fullerene aggregate that originate from molecular orbitals close to the LUMO orbital of PCBM,
we also account for the electronic states stemming from orbitals situated at around 1 eV above the LUMO orbital.
Our analysis reveals the importance of the space-separated states that inherit nonzero oscillator strengths from donor states
(bridge states and PACB states) and exhibit charge delocalization in ultrafast exciton dynamics.
Depending on the energy region of the exciton spectrum, the origin of these states is different.
In the low-energy region of the spectrum, bridge states are formed as a consequence of the resonant mixing among exciton (i.e., two-particle) states,
while in the opposite part of the spectrum the resonant mixing between single-electron states in the two materials brings about the formation of PACB states.

The resonant mixing has been suggested to be the key physical mechanism responsible for the presence
of separated charges on ultrafast time scales following the excitation of a D/A heterojunction.~\cite{nmat12-29,FD.163.377,PhysRevB.88.205304,jpcl.7.536,ADMA:ADMA201402294,jacs.136.2876,JPhysChemLett.7.4830}
Employing the model of reduced dimensionality and studying its subpicosecond dynamics on a fully quantum level, we reach similar conclusions,
and thus believe that our one-dimensional model is capable of describing the essential physics behind ultrafast interfacial processes.
Our one-dimensional model does not provide a detailed description of, e.g., the role of fullerene cluster size and packing in the ultrafast dynamics.
However, it takes into account the most important consequences of the aforementioned effects, i.e.,
the delocalization of electronic states and
the accessibility of delocalized states of space-separated charges from the states of donor excitons.~\cite{jacs.136.2876}
The effects of the dimensionality of the model become crucial on somewhat longer time and length scales.
Namely, on $\gtrsim 10$-ps time scales, the diffusion-controlled charge separation by incoherent hops throughout the respective materials takes place,~\cite{ncomms4-2334}
and one has to take into consideration all possible separation paths the electron and hole can follow, which can be done correctly only within a three-dimensional model
of the heterojunction.
The effects of electric polarization and screening at realistic interfaces are rather complex and strongly dependent on the details of the interface.~\cite{JPhysChemC.117.12981}
Here, however, we take a minimal model of the electron-hole interaction that reproduces the most important features of the energetics of exciton states
obtained from experimental data or from more sophisticated models of polarization and screening at interfaces.
In that sense, our model can be considered as an effective model whose parameters were adjusted to yield realistic energetics of relevant exciton states.

While the PACB states in our model enhance ultrafast charge separation by acting as additional interfacial photon-absorbing states,
the most important characteristic of the bridge states is not their direct accessibility from the ground state, but their good coupling with the manifold of donor states.
Therefore, for donor excitons, the bridge states act as gateways to the space-separated manifold,
so that the populations of low-lying space-separated states are built by progressive deexcitation within the space-separated manifold on a picosecond time scale following the excitation.

The ultrafast exciton dynamics strongly depends on the central frequency of the excitation.
While exciting well above the lowest donor state there are a number of photophysical pathways enabling subpicosecond exciton dissociation,
exciting at the lowest donor state the major part of generated excitons reside in this state and ultrafast exciton dissociation is not pronounced.
Stronger carrier-phonon interaction enhances phonon-mediated transitions from donor states to bridge states and is thus beneficial to exciton dissociation on ultrafast time scales.

Our results indicate that the number of space-separated charges that are present 1 ps after photoexcitation is rather small,
being typically less than 10\% of the number of excited electron-hole pairs (see, e.g., Figures~\ref{Fig:coh_incoh_tot_incoh_xd_cs_ct_xa} and~\ref{Fig:e_ph_infl_1}).
On the other hand, in most efficient solar cell devices internal quantum efficiencies (IQE) close to 100\% have been reported.
In light of an ongoing debate on the origin of high IQE and the time scale necessary for the charge separation process to occur,
our results indicate that longer time scales are needed to separate the charges.
Many of the photophysical pathways that we identify eventually lead to occupation of low-lying CT states
(e.g., in Figure~\ref{Fig:menagerie}, the pathway starting from initial donor excitons (black bolt)[$\rightarrow$(3)]$\rightarrow$(4)$\rightarrow$(5)$\rightarrow$(7)
or the pathway starting from initial PACB excitons (red bolt)[$\rightarrow$(1)]$\rightarrow$(2)[$\rightarrow$(3)]$\rightarrow$(4)$\rightarrow$(5)$\rightarrow$(7)).
We also find that on subpicosecond time scales a large portion of excitons remains in donor states.
Therefore, a mechanism that leads to escape of charges from low-lying CT states and donor states on longer time scales and consequently to high IQE needs to exist.
Several recent experimental~\cite{nmat13-63,PhysRevB.95.195301} and theoretical~\cite{ADMA:ADMA201304241,JPhysChemLett.7.4495,JPhysChemLett.8.2093}
studies have provided evidence that the separation of charges residing in these states is indeed possible.
Along these lines, our model could potentially be part of a multiscale model of the OPV devices,
as it yields the populations of different states at $\sim 1$ ps after photoexcitation.
The output of our model could then be used as input for a semiclassical model that would consider the charge separation and transport on a longer time scale.

\acknowledgments
We gratefully acknowledge the support by the Ministry of Education, Science and Technological
Development of the Republic of Serbia (Project No. ON171017) and
the European Commission under H2020 project VI-SEEM, Grant No. 675121,
as well as the contribution of the COST Action MP1406.
Numerical computations were performed on the PARADOX supercomputing
facility at the Scientific Computing Laboratory of the Institute of Physics Belgrade.
\newpage
\bibliography{refs-spoj}

\end{document}